\documentclass[preprint, amsmath, amssymb, aps, groupedaddress]{revtex4-1}
\usepackage{natbib}
\usepackage{graphicx, subfigure}
\usepackage{dcolumn}
\usepackage{bm}

\usepackage{indentfirst}
\usepackage{array, threeparttable}
\usepackage{multirow}
\usepackage{mathrsfs}
\usepackage{color}

\begin{document}
	
	
\title{A high order kinetic flow solver based on flux reconstruction framework}
	
	
\author{Ji Li}
\email[Ji Li: ]{leejearl@mail.nwpu.edu.cn}
\author{Chengwen Zhong}
\email[Corresponding author: Chengwen Zhong, ]{zhongcw@nwpu.edu.cn}
\author{Sha Liu}%
\email[Sha Liu: ]{shaliu@nwpu.edu.cn}

\affiliation{National Key Laboratory of Science and Technology on Aerodynamic Design and Research, Northwestern Polytechnical University, Xi'an, Shaanxi 710072, China.}
	
\date{\today}
	
\begin{abstract}
The goal of this paper is to develop a high order numerical method based on Kinetic Inviscid Flux (KIF) method and Flux Reconstruction (FR) framework. The KIF aims to find a balance between the excellent merits of Gas-Kinetic Scheme (GKS) and the lower computational costs. The idea of KIF can be viewed as an inviscid-viscous splitting version of the gas-kinetic scheme, and Shu and Ohwada have made the fundamental contribution. The combination of Totally Thermalized Transport (TTT) scheme and Kinetic Flux Vector Splitting (KFVS) method are achieved in KIF. Using a coefficient which is related to time step $\delta t$ and averaged collision time $\tau$, KIF can adjust the weights of TTT and KFVS flux in the simulation adaptively. By doing the inviscid-viscous splitting, KIF is very suitable and easy to integrate into the existing framework. The well understood FR framework is used widely for the advantages of robustness, economical costs and compactness. The combination of KIF and FR is originated by three motivations. The first purpose is to develop a high order method based on the gas kinetic theory. The second reason is to keep the advantages of GKS. The last aim is that the designed method should be more efficient. In present work, we use the KIF method to replace the Riemann flux solver applied in the interfaces of elements. The common solution at the interface is computed according to the gas kinetic theory, which makes the combination of KIF and FR scheme more reasonable and available. The accuracy and performance of present method are validated by several numerical cases. The Taylor-Green vortex problem has been used to verify its potential to simulate turbulent flows.
\end{abstract}
	
	
\maketitle
	
\clearpage

\section{Introduction}
The Gas-Kinetic Scheme (GKS) developed by Xu~\cite{xu2001gas,xu2005multidimensional} is based on the idea of Bhatnagar et al.~\cite{bhatnagar1954model}. In recent years, GKS is on the way to become the preferred numerical method in the fluid dynamics. Compared with traditional Navier-Stokes numerical method, GKS is of high spatial and temporal accuracy. The advantages of GKS have been recognized in the simulation of turbulent flows~\cite{li2010numerical, xiong2011numerical,FLD:FLD4239,li2016gas, cao_implicit_2019}, shock-boundary interaction, hypersonic flows~\cite{li2005application,xu2005multidimensional} and non-equilibrium simulations~\cite{liu2012multiple, zhu2010theoretical}. A series studies based on the GKS have been advanced, such as immersed boundary method~\cite{yuan2015immersed,YUAN2018417, AAMM-11-1177}, implicit temporal marching~\cite{li2014implicit}, and dual-time strategy~\cite{PhysRevE.95.053307} for unsteady flows.

In the field of Computational Fluid Dynamics (CFD), a numerical algorithm can be classified as a low or high order method according to the numerical accuracy approached. For the features of robustness and economical costs, the low order methods are still popular in the aeronautical industry. Compared with the low order method, the high order method is more accurate, which has the potential of providing more details of the flow fields~\cite{fujii2005progress,visbal2002use}. However, using the high order method in the real industry is still a challenge, which attracts the interests of many researchers from all over the world.  The development of high order GKS can be traced back to the study of Q. Li~\cite{li2010high}. J. Luo~\cite{luo2013high}, G. Zhou~\cite{zhou2017simplification}, L. Pan~\cite{pan2016efficient,pan2015third,pan2015compact}, X. Ji~\cite{ji_compact_2018,ji_hweno_2020}, F. Zhao~\cite{zhao_compact_2019} have made the great contribution to the high order gas-kinetic scheme.

Although high order gas-kinetic scheme has been studied quite well, the development of high order gas-kinetic scheme is never stopped. In present work, we focus on the combination of Kinetic Inviscid Flux (KIF) and Flux Reconstruction method (FR). The combination of KIF and FR is originated by three motivations. The first purpose is to develop a high order method based on the gas kinetic theory. The second reason is to keep the advantages of GKS. The last aim is that the designed method should be more efficient. In order to find a balance between the advantages of gas-kinetic scheme and lower computational costs, KIF is proposed by S. Liu\cite{Liu2020AMK}. The KIF scheme is a combination of Totally Thermalized Transport (TTT) scheme and Kinetic Flux Vector Splitting method (KFVS). The TTT scheme, which does not introduce extra artificial viscosity in smooth flow area, can approach the boundary layer accurately. TTT has the property similar to central schemes, so it also can not capture the discontinuity properly. The KFVS is a shock capturing scheme with good robustness. The combination is a good idea, which means that we use TTT scheme where the flow is smooth and use KFVS method where discontinuity exists. The kernel of KIF method is to adjust the weights of TTT and KFVS in the simulation automatically.

To develop a high order kinetic flux solver, it is critical to adopt the advantages from the traditional high order method based on the Navier-Stokes equation. Over the last 20 years, the high order numerical method is one of the research hotspots in the field of CFD. A great many of researchers have devoted their attentions to such a challenge, and a large numbers of high order numerical methods have been developed under the frameworks of finite volume method, finite difference method and finite element method et al. Some of the schemes are of notable features and have been used widely. For example, k-exact method~\cite{barth1990higher}, Essentially Non-Oscillatory (ENO) method
~\cite{abgrall1994essentially,durlofsky1992triangle,ollivier1997quasi,sonar1997construction}, Weighted ENO (WENO) method~\cite{liu1994weighted,shu1998essentially,hu1999weighted}, Discontinuous Galerkin (DG) method~\cite{cockburn2001runge}, radial basis function method~\cite{LIU20161096}, Compact Least-Squares (CLS) reconstruction method~\cite{WANG2016863, WANG2016883} and variational reconstruction method~\cite{wang_compact_2017}. An excellent review of the high order methods can be referred to the work of Z. Wang~\cite{wang2007high}.

Flux reconstruction method, first proposed by H. T. Huynh~\cite{Huynh2007, Huynh2009}, is aimed to be more popular in both of the research and real industry fields. The designed features of robustness, economical costs and compactness make it well understood and available. A particular FR scheme depends on three factors~\cite{vincent2011new},
namely the distribution of solution points, the Riemann flux solver applied at the interfaces, and the choice of the correction functions $\mathcal{G}$ and $\mathcal{H}$. It has been proved that the flux reconstruction method can recover the simplified DG and staggered grid scheme with specific factors, and the conservation also has been proved in Ref.~\cite{Huynh2007}. Based on the study of Jameson~\cite{jameson2010proof}, a class of energy stable flux reconstruction method was proposed by Vincent, Castonguay and Jameson (VCJH)~\cite{vincent2011new}. And then, the VCJH scheme was used for triangular elements~\cite{castonguay2012new}. Up to now, the VCJH correct function has played an important role in the FR framework.

In present work, the combination of KIF and FR is achieved by (a) replacing the Riemann solver applied on the interface of elements with KIF, (b) using the gas kinetic theory to compute the common solution on the interface, and (c) implementing the inviscid-viscous splitting strategy in the simulation. The present paper is organized as follows. In Sec.~\ref{fr-kif-method-introduction}, KIF method and the flux reconstruction framework are introduced. Several numerical tests are set up in the Sec.~\ref{cases}, and the numerical accuracy of present method is validated. The last section of paper is a short conclusion.

\section{Numerical method}\label{fr-kif-method-introduction}
The FR method, which takes advantages of DG and staggered grid scheme~\cite{KOPRIVA1996244, KOPRIVA1996475}, is first developed by H. T. Huynh~\cite{Huynh2007, Huynh2009}. It focuses on the features of robustness, economical costs and compactness. Benefiting from the merits of well understood and available, the flux reconstruction method has attracted a great many of attentions. In this section, the high order kinetic flux solver based on flux reconstruction framework will be introduced.

\subsection{Governing equation}
For one-dimensional problem, the Bhatnagar-Gross-Krook (BGK) model~\cite{bhatnagar1954model} in x-direction is
\begin{equation}\label{BGK_1D_x}
	f_t + uf_x = \frac{g-f}{\tau},
\end{equation}
where $u$ is the particle velocity, $f$ represents the gas distribution function, $g$ denotes the equilibrium state approached by $f$, and $\tau$ is related to the averaged collision time. The equilibrium state is known as a Maxwellian distribution reads
\begin{equation}\label{Maxwellian_1D_x}
	g = \rho (\frac{\lambda}{\pi})^{\frac{K+1}{2}}e^{-((u-U)^2+\xi^2)},
\end{equation}
where $\rho$ is the density, $U$ is the macroscopic velocity. $\lambda$, which reads $\lambda = m/(2kT)$, is related to the temperature $T$ of gas, $m$ represents the molecular mass, and $k$ denotes the Boltzmann constant. The total number of degrees of freedom $K$ in $\xi$ equals to $(5-3\gamma)/(\gamma-1) + 2$, $\gamma$ is the ratio of specific heat, and $\bm{\xi}^2 = \sum_{i=1}^{K}\xi_i^2$.

According to the kinetic theory of gases, both the distribution function $f$ and the equilibrium state $g$ are functions of particle velocities $u$, space $x$ and time $t$. Taking the moments of distribution function $f$, the macroscopic conservative variable $\bm{w}$ can be obtained as follow
\begin{equation}\label{macroVar}
	\bm{w} = \left(
	\begin{array}{c}
		\rho \\
		\rho U \\
		E
	\end{array}
	\right) = \int {\bm{\psi}fd\Xi},\quad \bm{\psi} = \left(1, u, \frac{1}{2}\left(u^2 + \bm{\xi}^2 \right) \right)^T,
\end{equation}
where $d\Xi = du \left(\prod\limits_{i=1}^{K}d\xi_i \right)$. In order to obtain the spatial discretization in the flux reconstruction framework, we take moments of
 $\bm{\psi}$ in Eq.~\eqref{BGK_1D_x} and integrate it with $d\Xi$ in phase space,
\begin{equation}\label{fvm1}
	\int \left( f_t + uf_x\right)\bm{\psi}d\Xi = -\int \frac{f-g}{\tau}\bm{\psi}d\Xi.
\end{equation}
Using the compatibility condition
\begin{equation}\label{compatibilityCondition}
	\int \frac{g-f}{\tau}\bm{\psi}d\Xi = 0.
\end{equation}
We can get the following formula
\begin{equation}\label{FR_1D_adv}
	\bm{w}_t + \bm{G}_x = 0, \quad \bm{G}=\int uf\bm{\psi d}\Xi,
\end{equation}
where $\bm{G}$ is the flux corresponding to conservative variables $\bm{w}$ along the x-direction. Then, we can solve the Eq.~\eqref{FR_1D_adv} within the flux reconstruction framework, and the flux $\bm{G}$ can be computed using KIF.

\subsection{Kinetic Inviscid Flux}\label{kif-method-introduction}
The Riemann flux solver employed at the interfaces is one of the three critical factors of flux reconstruction method. In the present work, we implement the kinetic inviscid flux to determine the common flux at the interface of elements. The motivation of our work is to reach a compromise between good performance of gas-kinetic scheme and lower computational costs. The KIF is a kind combination of TTT scheme and KFVS method.

TTT scheme has been discussed by Xu in Ref.~\cite{1993PhDTxukun}. The first step of TTT scheme is to get the Maxwell distribution
on both sides of the surface,

\begin{equation}\label{Maxwellian_1D_lr}
g_l = [\rho (\frac{\lambda}{\pi})^{\frac{K+1}{2}}e^{-((u-U)^2+\xi^2)}]_l, \quad g_r = [\rho (\frac{\lambda}{\pi})^{\frac{K+1}{2}}e^{-((u-U)^2+\xi^2)}]_r.
\end{equation}
The subscripts $l$ and $r$ represent the state based on the macroscopic variables from left and right sides of interface respectively.
The second step is to make particles crossing
the interface collide sufficiently. The new Maxwell distribution $g_0$ is assumed to have the following form
\begin{equation}\label{non-eq2}
g_0 = [\rho (\frac{\lambda}{\pi})^{\frac{K+1}{2}}e^{-((u-U)^2+\xi^2)}]_0, \quad \bm{w}_0=\int g_0\psi d\Xi = \int  ((1-H(x))g_l + H(x)g_r) \psi d\Xi,
\end{equation}
where $H(x)$ is the Heaviside function
\begin{equation}\label{heaviside}
H(x) = \left \{
\begin{array}{cc}
0, & x < 0, \\
1, & x \geq 0.
\end{array}
\right.
\end{equation}
Finally, the flux of TTT scheme reads
\begin{equation}\label{flux_ttt}
F_{TTT} = \int ug_0\psi d\Xi.
\end{equation}
The TTT scheme has the similar property to central scheme, which does not introduce extra artificial viscosity in smooth flow area and can capture the boundary layer accurately. However, it can not deal with shock wave, because of lacking artificial viscosity.

Equilibrium Flux Method (EFM)~\cite{Pullin1980}  and KFVS~\cite{mandal1994kinetic} are similar, and we just call it KFVS here. KFVS is another kinetic scheme and is a shock capturing method. After getting the Maxwell distribution beside the interface using Eq.~\eqref{Maxwellian_1D_lr}, KFVS gives flux by calculating particles across the interface as
\begin{equation}\label{flux_kfvs}
F_{KFVS} = \int_{u>0} ug_l\psi d\Xi + \int_{u<0} ug_r\psi d\Xi.
\end{equation}
KFVS has the properties as good robustness and positivity-preserving~\cite{tao_gas-kinetic_1999}, which make the scheme suitable for capturing the discontinuity. However, it introduces enormous artificial viscosity, and the essential problem, which is analyzed in Ref.~\cite{1993PhDTxukun}, is that the equation solved at interface is the collisionless Boltzmann equation.

Combination is a good idea, which means that we use TTT in the smooth flow area and KFVS in the flow field where shock wave exists.
The idea of KIF can be viewed as an inviscid-viscous splitting version of the gas-kinetic scheme, and Shu~\cite{sun2016a} and Ohwada~\cite{ohwada2018a} have made the fundamental contribution. The most significant difference between KIF~\cite{Liu2020AMK} and the works of Shu and Ohwada is the weight of TTT and KFVS. Ref.~\cite{Liu2020AMK} adopted the philosophy of direct modeling~\cite{xu2015direct}, and constructed two kinds of KIF method (namely KIF1 method from
GKS strategy~\cite{xu2001gas} and KIF2 from DUGKS strategy~\cite{guo2013discrete}). In present work, we adopt the KIF1 in the simulations. The KIF1 can be expressed as

\begin{equation}\label{kif1}
\bm{F} = \{\frac{\tau}{\delta t}(1-e^{-\delta t/\tau})\}\bm{F}_{KFVS} + \{1-\frac{\tau}{\delta t}(1-e^{-\delta t/\tau})\}\bm{F}_{TTT},
\end{equation}

\begin{equation}\label{kif_r}
\delta t = r\tau = \frac{\tau}{\max \limits_{\Omega}\left[\frac{|p_l-p_r|}{|p_l+p_r|}, \max(Ma_l, Ma_r)\right]}.
\end{equation}
$\delta t$ is the observation time scale and is measured in mean collision time (or the relaxation time $\tau$) in discontinuities. Since KIF is an inviscid-viscous splitting version of GKS, the viscous fluxes can be computed using the traditional central viscosity scheme. The details of KIF1 can be referred to Ref.~\cite{Liu2020AMK}.

In another point of view, KIF is a kind of balance between kinetic scheme and traditional macroscopic numerical method. In recent years, kinetic schemes have a significant development~\cite{qu2007alternative, xu_unified_2010, guo2013discrete, wu2015a, liu2020unified}, which mainly aim at nonequilibrium flow. With a view at equilibrium state, KIF replaces the complicated nonequilibrium part by traditional central viscosity scheme. One motivation is to find a balance between advantages and efficiency, while another is to be suitable and easy to integrate into the existing framework.

\subsection{Spatial discretization in flux reconstruction framework}\label{FR_1D_introduction}
In this section, the FR framework used to solve Eq.~\eqref{FR_1D_adv} is introduced. For the one-dimensional problem,
the computational domain $\Omega$ can be divided into $N$ subdomains
\begin{equation} \label{spatial_discretizalition_1D}
	\Omega = \{\Omega_i|i = 0, 1, \cdots, N-1\}, \quad \Omega_i = [x_i, x_{i+1}],  \quad x_0 < x_1 < \cdots < x_N.
\end{equation}
Within the element $\Omega_i$, the solution points are set as $x_{i,k}$ ($k=0,1,2,\cdots,P$). It is obvious that the number of solution points within a standard elements is $P+1$, and $P$ is related with the accuracy order of the numerical method. The set $x_{i,k}$ can be chosen as  Gauss, Radau, Lobatto or equidistant points. It has been proved in Refs.~\cite{Huynh2007,Huynh2009} by H. T. Huynh that Fourier stability and accuracy analysis of FR framework are independent of the type of solution points.

Since dealing with every elements $\Omega_i$ is very tedious, all the element $\Omega_i$ should be mapped into the same standard element $\Omega_s = \{\xi | \xi \in [-1, 1]\}$ to simplify the implementation of the algorithm. The mapping function $\theta(\xi)$ can be expressed as
\begin{equation}\label{mapping_1D}
	x = \theta_i(\xi) = \left(\frac{1-\xi}{2}\right)x_i + \left(\frac{1+\xi}{2}\right)x_{i+1}.
\end{equation}
In order to be consistent with the existed literature of FR framework, the denotation of flux $\bm{G}$ in Eq.~\eqref{FR_1D_adv} is replaced by $\bm{f}$. With the mapping expressed as Eq.~\eqref{mapping_1D}, the evolution of macroscopic variables $\bm{w}$ within each $\Omega_i$ can be transformed as the Eq.~\eqref{FR_1D_adv_s} within the standard element.
\begin{equation}\label{FR_1D_adv_s}
	\bm{\hat{w}}_t + \frac{1}{J_n}\bm{\hat{f}}_\xi = 0,
\end{equation}
where
\begin{equation}\label{w_s}
	\bm{\hat{w}} = \bm{w}(\theta_i(\xi),t) \quad in \quad \Omega_i,
\end{equation}

\begin{equation}\label{f_s}
	\hat{\bm{f}} = \bm{f}(\theta_i(\xi),t) \quad in \quad \Omega_i,
\end{equation}

\begin{equation}\label{Jn}
	J_n = \frac{\partial x}{\partial \xi} \quad in \quad \Omega_i.
\end{equation}

The FR framework for solving the Eq.~\eqref{FR_1D_adv_s} within the standard element $\Omega_s$ consists of seven subsequent steps. In the first step, the solution polynomial $\bm{\hat{w}}_i^{\delta}(x)$ can be obtained through the macroscopic variable $\bm{\hat{w}}_{i,k}$ at the solution points $\xi_{k}$,
 \begin{equation}\label{solution_poly}
 	\bm{\hat{w}}_i^{\delta}(x) = \sum_{k=0}^{k=P} {\bm{\hat{w}}_{i,k}\phi_{k}(\xi_k)},
 \end{equation}
 where the symbol $\delta$ denotes the solution polynomial is always discontinuous at the element interface. $\phi_{k}(\xi_k)$ is the 1D Lagrange polynomial equal to $1$ at the kth solution point and $0$ at the others,
 \begin{equation}\label{Lagrange_Poly_1D}
 	\phi_{k}(\xi_k) = \prod\limits_{l=0, l \neq k}^{P}{\frac{\xi - \xi_l}{\xi_k-\xi_l}}.
 \end{equation}

 \begin{figure}
 	\centering
 	\includegraphics[width=0.4 \textwidth]{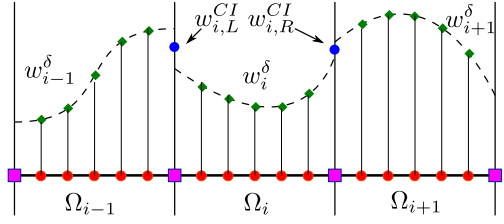}
 	\caption{\label{fig:solution_poly} The discontinuous solution polynomials and interface common solutions within elements $\Omega_{i-1}$, $\Omega_i$ and $\Omega_{i+1}$.}
 \end{figure}

 Fig.~\ref{fig:solution_poly} shows the solution polynomials at element $\Omega_i$ and the neighbors in the physical space. Take the interface $x_{i+1/2}$ as an example, $\bm{w}_i^{\delta}(x_{x+1/2})$ and $\bm{w}_{i+1}^{\delta}(x_{x+1/2})$ represent the macroscopic variables at the interface from $\Omega_i$ (left) and $\Omega_{i+1}$ (right) respectively. Generally speaking, $\bm{w}_i^{\delta}(x_{x+1/2})$ and $\bm{w}_{i+1}^{\delta}(x_{x+1/2})$ are not equal. Since $\xi$ belongs to the interval
 $[-1, 1]$ in the standard element $\Omega_s$, $\bm{w}_i^{\delta}(x_{x+1/2})$ and $\bm{w}_{i+1}^{\delta}(x_{x+1/2})$ equal to  $\bm{\hat{w}^{\delta}}_i(1)$ and $\bm{\hat{w}^{\delta}}_{i+1}(-1)$ respectively. It is natural that $\bm{\hat{w}^{\delta}}_i(1)$ does not equal to $\bm{\hat{w}}_{i+1}^{\delta}(-1)$ in most of cases, which is the ``Jump" or ``Discontinuous" at the boundaries of element.

 In the second step, it is turn to determine the common solution $\bm{\hat{w}}^{CI}$ at the boundaries of standard element, i.e. $\xi=\pm 1$. The common solution $\bm{\hat{w}}^{CI}$ at interface is used to make the solution within the standard element to feel the effect of the boundaries, so the superscript $C$ also  has the meanings ``Corrected" and ``Continuous". In present scheme, the common solution $\bm{\hat{w}}^{CI}$
 is computed using the following expression
 \begin{equation}\label{eq:w_c}
	\bm{\hat{w}}_{i+1/2}^{CI} = \int_{u>0} g_l\bm{\psi} d\Xi
+ \int_{u<0} g_r\bm{\psi} d\Xi,
 \end{equation}
where $g_l$ and $g_r$, which are corresponding to $\bm{\hat{w}}_i(1)$ and $\bm{\hat{w}}_{i+1}(-1)$, are the Maxwellian distributions at the left and right sides of an interface. $\bm{\hat{w}}_i^{\delta}(1)$ and $\bm{\hat{w}}_{i+1}^{\delta}(-1)$ can be obtained easily using Eq.~\eqref{solution_poly}. The demonstration of
$\bm{\hat{w}}^{CI}$ is shown in the Fig.~\ref{fig:solution_poly}.

\begin{figure}[!htp]
	\centering
	\subfigure[]{
		\includegraphics[width=0.4 \textwidth]{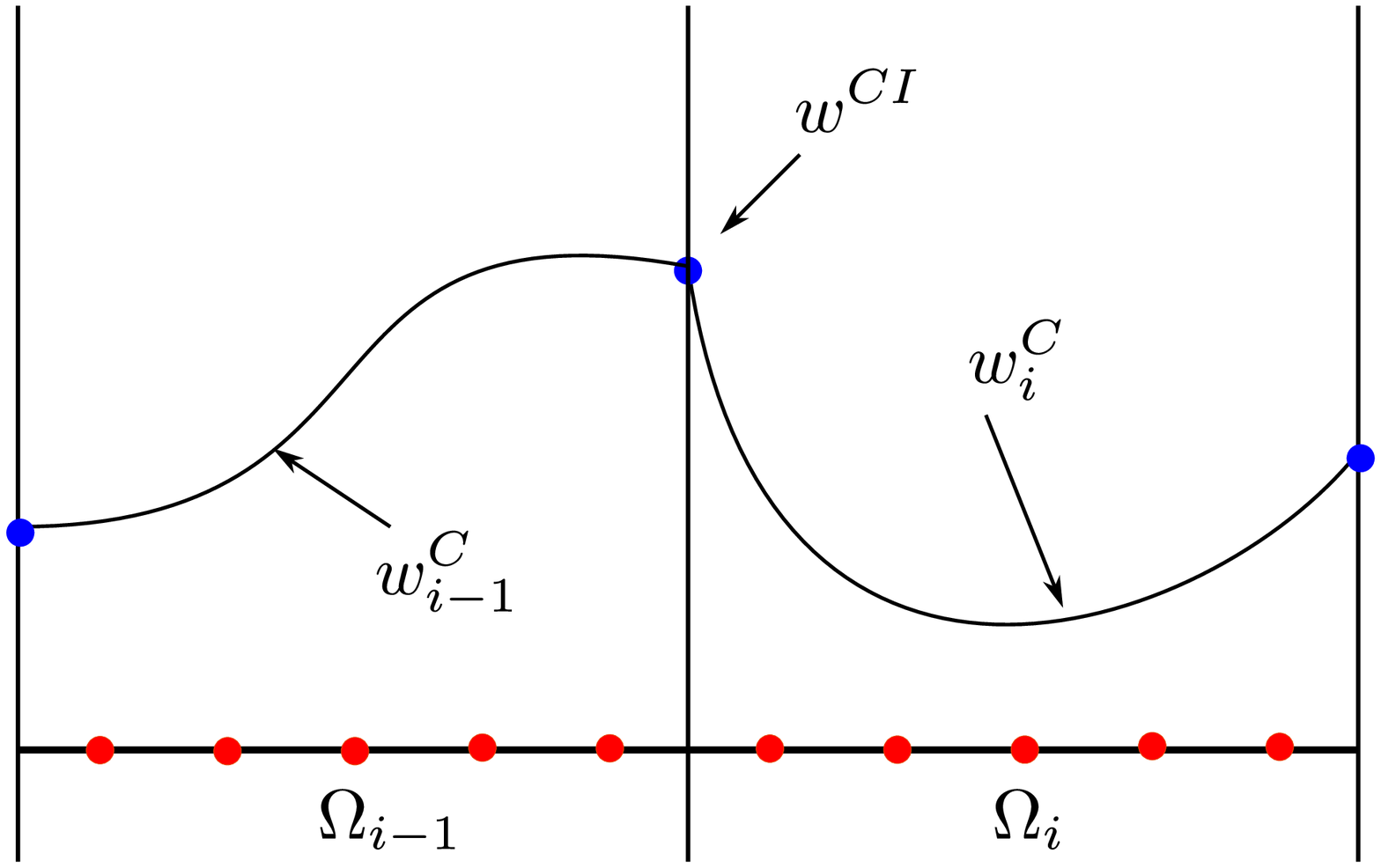}
		\label{fig:solution_correction:0}
	}
	\subfigure[]{
		\includegraphics[width=0.4 \textwidth]{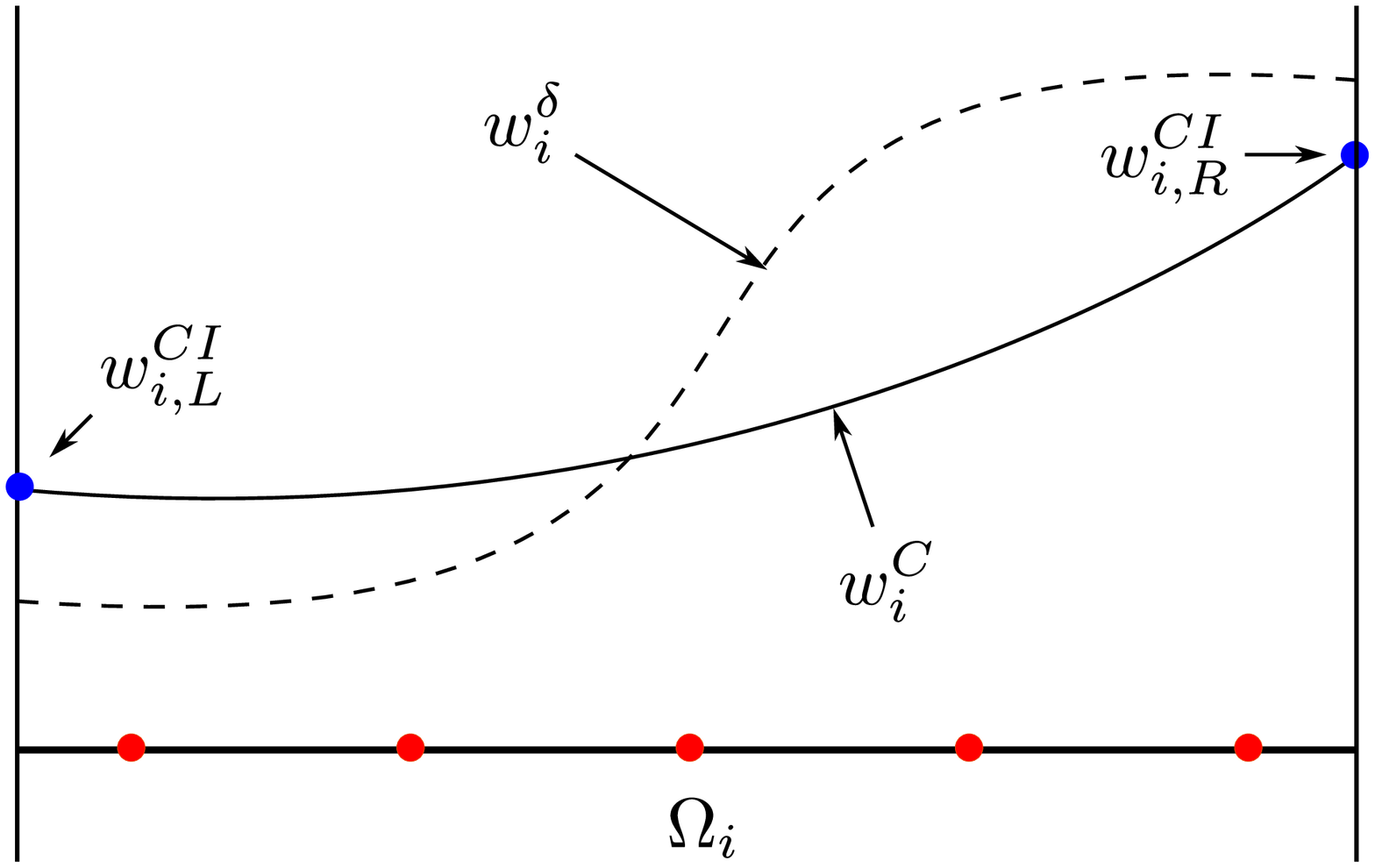}
		\label{fig:solution_correction:1}
	}
	\caption{The demonstration of solution correction: (a) Corrected solution beside the interface and
		(b) The solution correction within element $\Omega_i$.}
	\label{fig:solution_correction}
\end{figure}

It must be noticed that if the jumps at boundaries of element are ignored, the solution within the element can not
feel the effect of the boundaries and the evolution of scheme must get an erroneous result. Thus, the third step is to construct the corrected (or continuous) solution polynomial in the standard element. As shown in Fig.~\ref{fig:solution_poly}, the common solutions at the two end-points of the element of $\Omega_i$ are $\bm{\hat{w}}_{i,L}^{CI}$ and $\bm{\hat{w}}_{i,R}^{CI}$ respectively. The corrected solution polynomial within the element is named as $\bm{\hat{w}}_i^{C}(\xi)$, and must has features as
\begin{equation}\label{common_solution_features_1}
	\bm{\hat{w}}_i^{C}(-1) = \bm{\hat{w}}_{i-1/2}^{CI} = \bm{\hat{w}}_{i,L}^{CI}, \quad
	\bm{\hat{w}}_i^{C}(1) = \bm{\hat{w}}_{i+1/2}^{CI} = \bm{\hat{w}}_{i,R}^{CI}.
\end{equation}
The corrected solution polynomial $\bm{\hat{w}}_i^{C}(\xi)$ is assumed to have the following form
\begin{equation}\label{common_solution_poly}
	\bm{\hat{w}}_i^{C}(\xi) = \bm{\hat{w}}_i^{\delta}(\xi)
	+ (\bm{\hat{w}}_i^{C}(-1) - \bm{\hat{w}}_i^{\delta}(-1))\mathcal{G}_L(\xi)
	+ (\bm{\hat{w}}_i^{C}(1) - \bm{\hat{w}}_i^{\delta}(1))\mathcal{G}_R(\xi).
\end{equation}
$\mathcal{G}_L(\xi)$ and $\mathcal{G}_R(\xi)$ are the correct functions related with the left and the right end-points of the element, and $\mathcal{G}_L(\xi)$ and $\mathcal{G}_R(\xi)$ should satisfy the following conditions
\begin{equation}\label{correction_f_g}
	\mathcal{G}_L(-1) = 1, \quad \mathcal{G}_L(1) = 0, \quad
	\mathcal{G}_R(-1) = 0, \quad \mathcal{G}_R(1) = 1.
\end{equation}
The correct function is one of the most important factor of the FR Framework. More details and introductions of correct function can be referred to the works of H. T. Huynh\cite{Huynh2007, Huynh2009}. Now that the corrected solution polynomial is computed, the corrected solution derivatives polynomial $\bm{\hat{w}}_{i,\xi}^{C}(\xi)$ can be
obtained directly using the Eq.~\eqref{common_solution_poly}. The correction procedure can be seen in Fig.~\ref{fig:solution_correction} briefly.

The fourth step is to compute the flux $\hat{\bm{f}}_{i,k}$ at solution points. The flux at the solutions point can be evaluated by the macroscopic variables $\bm{\hat{w}_{i,k}}$ and the corrected derivatives $\bm{\hat{w}}_{i,\xi}^{C}(\xi_k)$. The flow is considered as continuous within the element, so the flux solver for smooth flow is used at the solution points. After the fluxes at solution points have been computed, the flux polynomial within the element can be obtained,
\begin{equation}\label{flux_poly}
\bm{\hat{f}}_i^\delta(x) = \sum_{k=0}^{k=P} {\bm{\hat{f}}_{i,k}\phi_{k}(\xi_k)}.
\end{equation}
$\bm{\hat{f}}_i^\delta$ means that the flux polynomial is always discontinuous at the boundaries of elements.

The fifth step focuses on the common flux at the two-end points of element. The fluxes within the whole computational domain are assumed as a piecewise function, which has the form as Eq.~\eqref{flux_poly} in each individual element. The fluxes across the boundaries of elements are always discontinuous. To make the fluxes within element feel the effect of boundaries, it is very important to compute the continuous (or common) fluxes at the interface to get the accurate results.

The common fluxes at boundaries of elements is another critical factor of the FR framework. The discontinuity is considered in the computation of common fluxes, and different Riemann solvers are used according to the numerical method. In our present work, the KIF method is applied to solve the fluxes across the interface.

\begin{figure}[!htp]
	\centering
	\subfigure[]{
		\includegraphics[width=0.4 \textwidth]{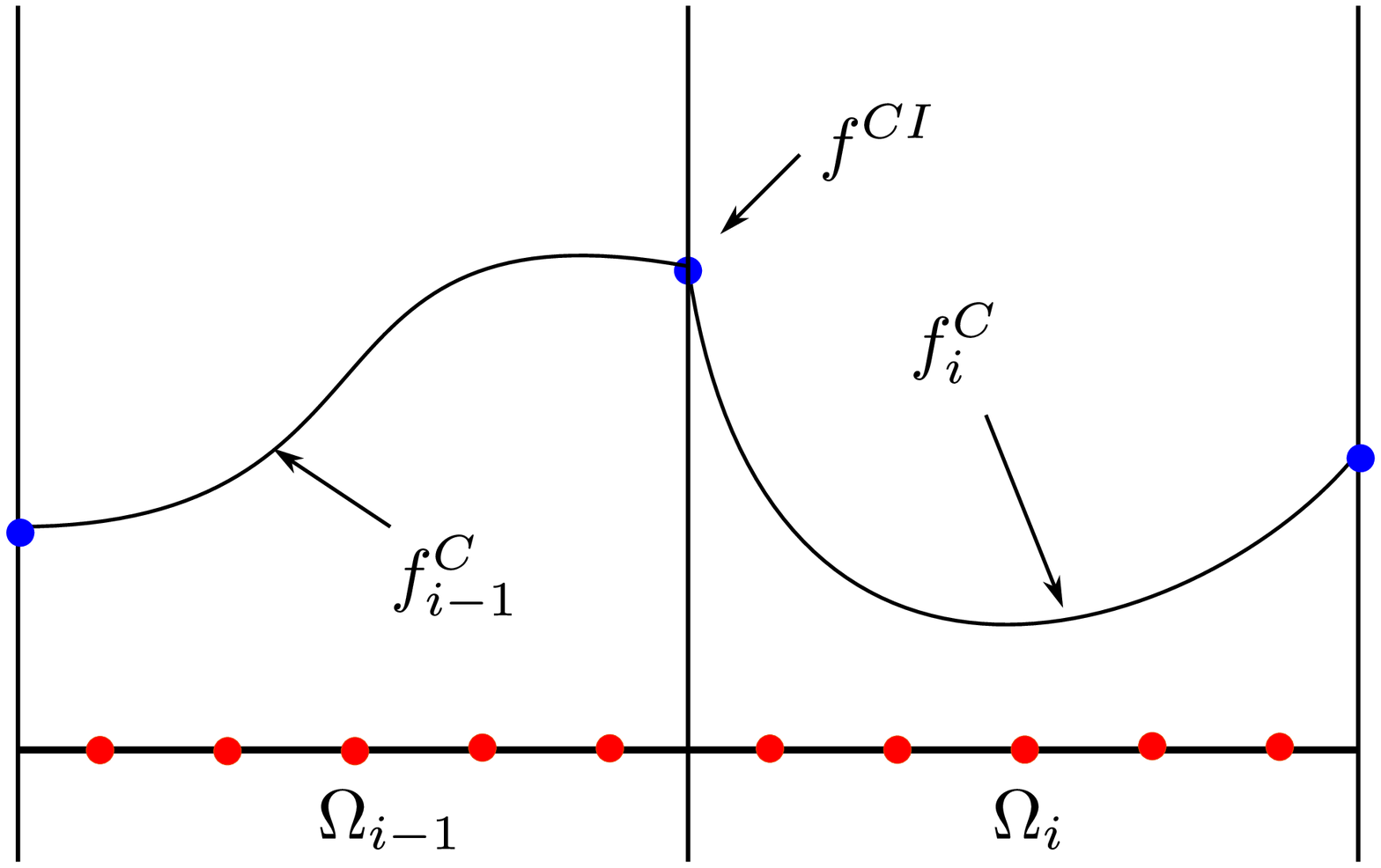}
		\label{fig:flux_correction:0}
	}
	\subfigure[]{
		\includegraphics[width=0.4 \textwidth]{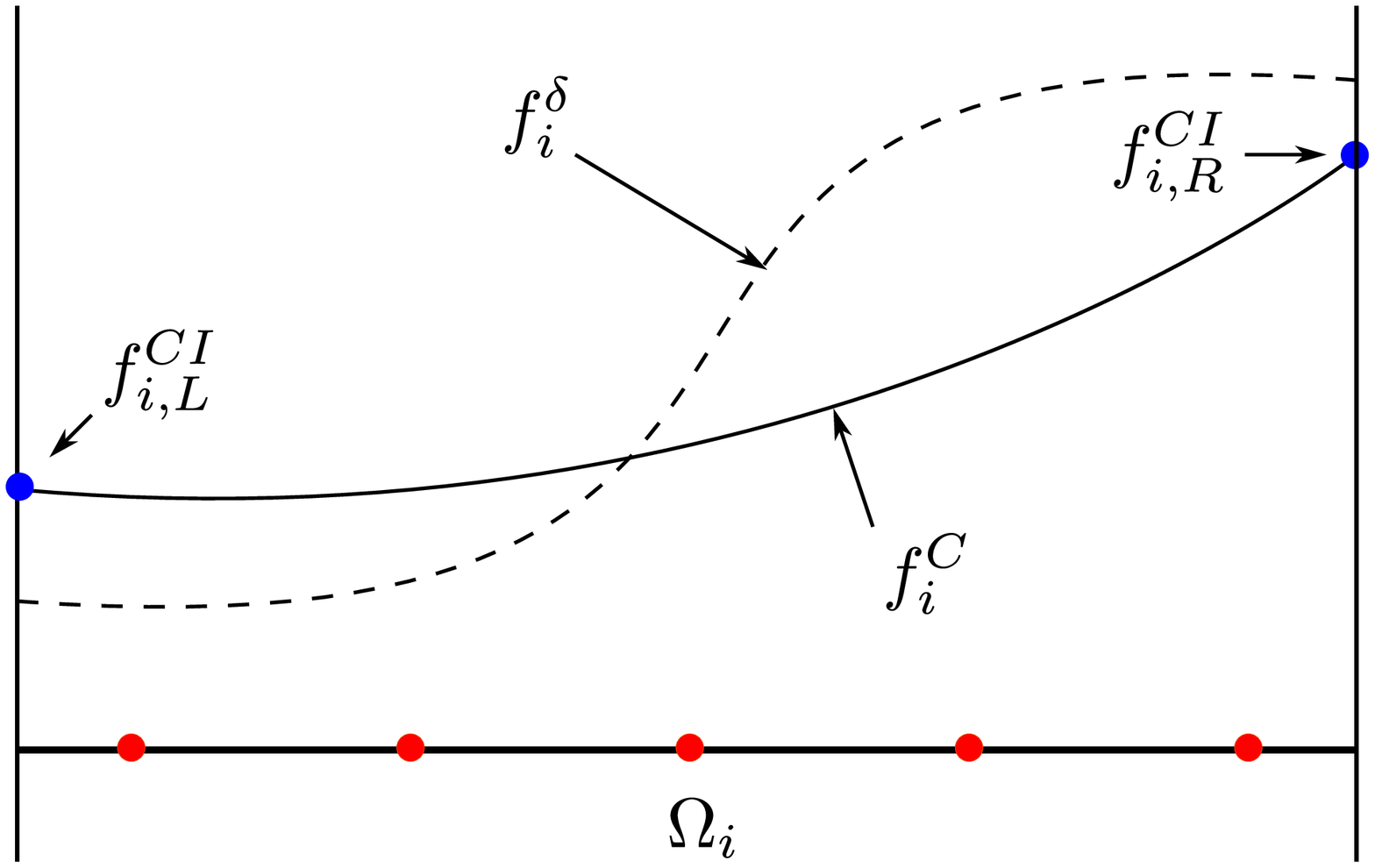}
		\label{fig:flux_correction:1}
	}
	\caption{The demonstration of flux correction: (a) Corrected flux beside the interface and (b) The flux correction within element $\Omega_i$.}
	\label{fig:flux_correction}
\end{figure}

The sixth step is to correct the fluxes using the common fluxes at the interfaces. The procedure of flux correction is very similar to solution correction, and the Fig.~\ref{fig:flux_correction} exhibits the procedure primitively. The corrected flux polynomial reads
\begin{equation}\label{corrected_flux_poly}
	\bm{\hat{f}}_i^{C}(\xi) = \bm{\hat{f}}_i^{\delta}(\xi)
		+ (\bm{\hat{f}}_i^{C}(-1) - \bm{\hat{f}}_i^{\delta}(-1))\mathcal{H}_L(\xi)
		+ (\bm{\hat{f}}_i^{C}(1) - \bm{\hat{f}}_i^{\delta}(1))\mathcal{H}_R(\xi),
\end{equation}
where $\mathcal{H}$ is the correction function, which is similar to the $\mathcal{G}$ used in the correction procedure of solution polynomial, and $\mathcal{H}_L(\xi)$ and $\mathcal{H}_R(\xi)$ should satisfy the following conditions
\begin{equation}\label{correction_f_h}
\mathcal{H}_L(-1) = 1, \quad \mathcal{H}_L(1) = 0, \quad
\mathcal{H}_R(-1) = 0, \quad \mathcal{H}_R(1) = 1.
\end{equation}

The final step is to compute the divergence of the corrected fluxes at the solution points. Since the corrected flux polynomial is expressed as Eq.~\eqref{corrected_flux_poly}, the divergence of corrected flux can be obtained directly. By now, all the preconditions of  using Eq.~\eqref{FR_1D_adv_s} to update the macroscopic variables at the solution points are completed for 1D advection problem.

The correction function is critical for the flux reconstruction method. $\mathcal{G}$ and $\mathcal{H}$ have a great effect on the accuracy and stability. The VCJH scheme developed by Vincent\cite{vincent2011new} is used in our work. The VCJH scheme has been proved as an energy stable scheme, and it can be recovered to a particular existing scheme, such as  nodal DG and Spectral Difference (SD) methods, with a specific parameter.

In terms of time integration, an explicit adaptive Runge-Kutta 45 (RK45) method is used in present work.

\subsection{Extension to multidimensional problem}
Extension to quadrilateral and hexahedral elements are straight forward~\cite{Huynh2007, Huynh2009}. For triangles~\cite{castonguay2012new, williams_energy_2013} and tetrahedra~\cite{williams2014energy}, they are more complicated in algorithm and implementation. But, the procedure is analogous to FR in one dimension.

\subsection{Shock capturing method}
The robust shock capturing is a main difficulty for high-order FE-type CFD method. In the vicinity of discontinuities, the smooth indicator~\cite{Persson2006-112, Persson2013-3061, Jameson_6.2014-2688} is used to detect the discontinuity. Once the shock has been sensed, the shock capturing method is applied on the elements. In the present work, we follow the idea of Jameson~\cite{Jameson_6.2014-2688}. We use two parameters $s_0$ and $\kappa$ to decide whether the shock capturing method should be applied. The determination of values $s_0$ and $\kappa$ can be referred to works~\cite{WangZJLLAV_2015, VANDENHOECK20191}. In the present paper, we set $s_0+\kappa$ around the value $0.01$.
We find this setting can keep the scheme robust and accurate.

\section{Numerical test cases}\label{cases}
In this section, numerical tests are set up for the validation of present method. The accuracy order, shock capturing method, viscous flow problem and various boundary conditions are all validated in the section. Finally, the potential of present scheme to simulate the turbulent flow is verified in the Taylor-Green Vortex problem.

Our algorithmic code is deployed on the HiFiLES open-source platform~\cite{HiFiLES2014}, thanks for their great works. It should also be noticed that the polynomial order $p=3$ is used in this section. Several one-dimensional problems are simulated using multidimensional code in present paper. The upper and bottom bounds of the computational domain are treated as periodic boundaries in these cases.

\subsection{Accuracy tests}
In this case, the advection of density perturbation problem~\cite{li2010high} is presented to validate the accuracy of our method on Cartesian grid. The initial condition is given as
\begin{equation}\label{InitDenPerturbation}
\rho(x) = 1+0.2sin(\pi x), \quad u(x) = 1, \quad v(x) = 0, \quad p(x) = 1,
\end{equation}
and the analytic solution at the time $t$ can be expressed as
\begin{equation}\label{SolDenPerturbation}
\rho(x, t) = 1+0.2sin(\pi (x-t)), \quad u(x, t) = 1, \quad v(x, t) = 0, \quad p(x, t) = 1.
\end{equation}
The case is a one-dimensional problem, and we simulate it using a two-dimensional solver on the Cartesian grid. The computational domain is
\begin{equation}
\{(x,y)| x \in [0,1], y \in [0, 4h]\},
\end{equation}
where the length scale $h$ equals to $1/N$. $N$ is the number of elements along the x-direction. Table~\ref{table:advden:1D} gives the $L_2$ normal of density distribution. The numerical results are obtained at $t = 2$, and the time step is set as $\Delta t = 0.05/N$. It can be concluded from the Table~\ref{table:advden:1D} that the accuracy order is reached quite well.
	\begin{table}[!htp]
	\centering
	\caption{\label{table:advden:1D} $L_2$ normal of density for the advection of density problem}
	\begin{threeparttable}
		\begin{tabular}{p{10pt} p{100pt}< {\centering} p{100pt}< {\centering}}
			\hline
			\hline
			$N$ & $L_2$  & order  \\
			\hline
			$10$ & $2.711254E-05$ & $-$  \\
			
			$20$ & $1.004359E-06$ & $4.754614$  \\
			
			$40$ & $5.789748E-08$ & $4.116630$  \\
			
			$80$ & $3.339470E-09$ & $4.115809$  \\
			
			$160$ & $2.089252E-10$ & $3.998561$  \\
			
			$320$ & $1.310146E-11$ & $3.995187$  \\
			
			$640$ & $8.208407E-13$ & $3.996481$  \\
			\lasthline
		\end{tabular}
	\end{threeparttable}
\end{table}

The second case is the isentropic vortex problem, which is a two-dimensional problem always used to validate the accuracy of high order method. The computational domain is a $[-5, 5]\times[-5,5]$ square. The periodic boundary is applied on the four bounds of the square. The diagonal uniform form flow, $\left(\rho, u, v, p\right)= \left(1, 1, 1, 1\right)$, is initialed in the flow field. Then, a small perturbation is added to the center of the square,

\begin{equation}\label{isentropic_vortex:1}
(\delta u, \delta v) = \frac{\epsilon}{2\pi}e^{(0.5(1-r^2))}(-\bar{y},\bar{x}),
\end{equation}
\begin{equation}\label{isentropic_vortex:2}
\delta T = -\frac{(\gamma-1)\epsilon^2}{8\gamma \pi^2}e^{(1-r^2)},\quad \delta S =0.
\end{equation}
where
\begin{equation*}
(\bar{x},\bar{y})=(x,y).\quad r^2=(\bar{x}^2+\bar{y}^2).
\end{equation*}
Since the vortex is moved along the diagonal line with the time marching in the case, the numerical results is obtained at $t=10$. The vortex is just back to the origin position at the moment. The triangular grid, which is similar to the grid shown in Fig.~\ref{fig:cavity:grid_tri}, is used in the simulation. The time step is set as $\Delta t = 0.01/N$. $N$ is the number of elements on the bound. Table~\ref{table:2D:isentropic_vortex} gives the $L_2$ normal of density. The designed accuracy order can be clearly seen in the table.

\begin{table}[!htp]
	\centering
	\caption{\label{table:2D:isentropic_vortex} $L2$ normal of density for the isentropic vortex problem.}
	\begin{threeparttable}
		\begin{tabular}{p{10pt} p{100pt}< {\centering} p{100pt}< {\centering}}
			\hline
			\hline
			$N$ & $L_2$  & order  \\
			\hline
			$10$ & $4.342904E-03$ & $-$  \\
			
			$20$ & $2.410620E-04$ & $4.171184$  \\
			
			$40$ & $1.488880E-05$ & $4.017105$  \\
			
			$80$ & $9.680808E-07$ & $3.942956$  \\
			
			$160$ & $7.091912E-08$ & $3.770881$  \\			
			
			\lasthline
		\end{tabular}
	\end{threeparttable}
\end{table}

\subsection{One dimensional Riemann problem}
\begin{figure}[!htp]
	\centering
	\subfigure[]{
		\includegraphics[width=0.4 \textwidth]{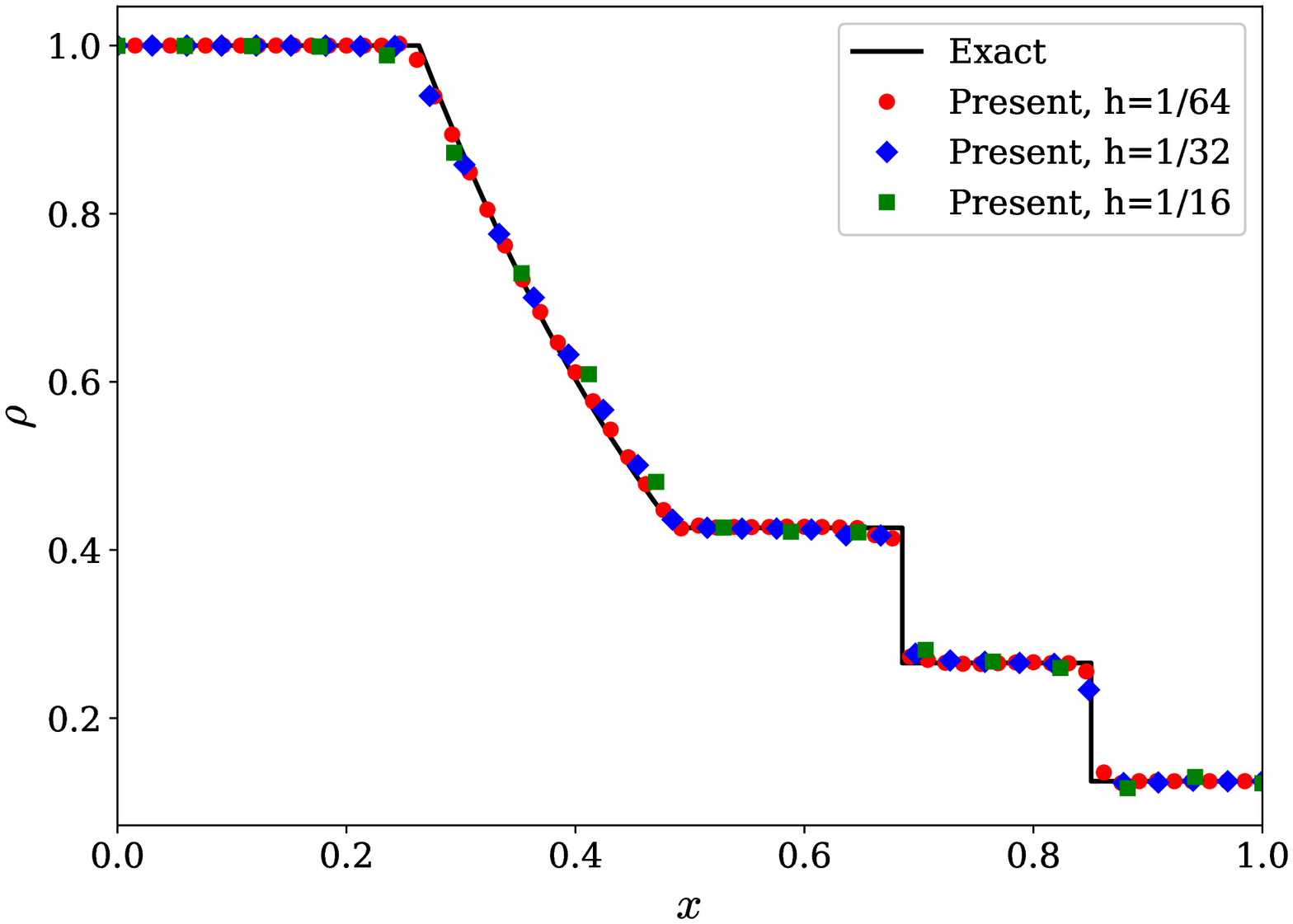}
		\label{fig:sod:rho:h_refine}
	}
	\subfigure[]{
		\includegraphics[width=0.4 \textwidth]{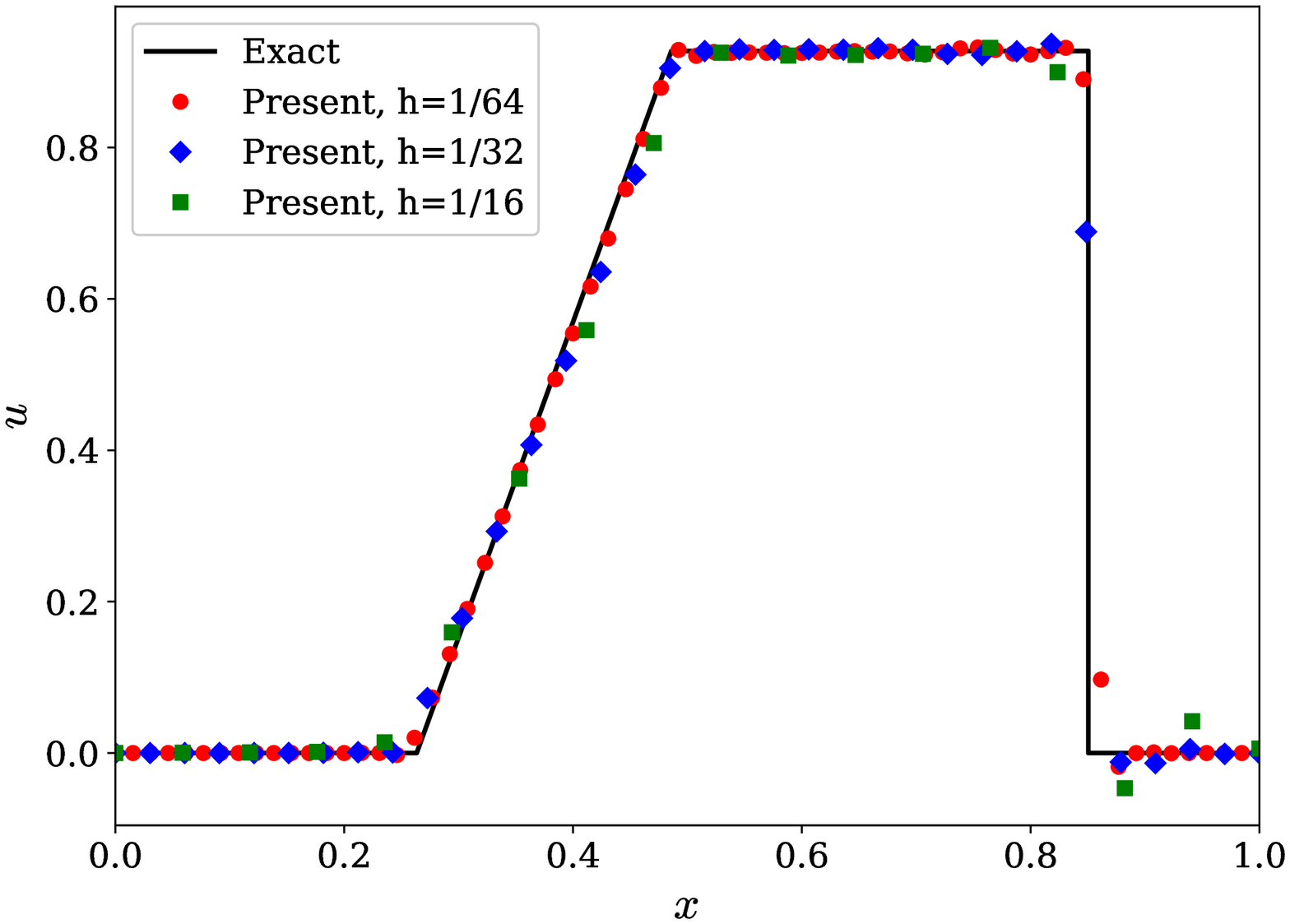}
		\label{fig:sod:u:h_refine}
	}
	\subfigure[]{
		\includegraphics[width=0.4 \textwidth]{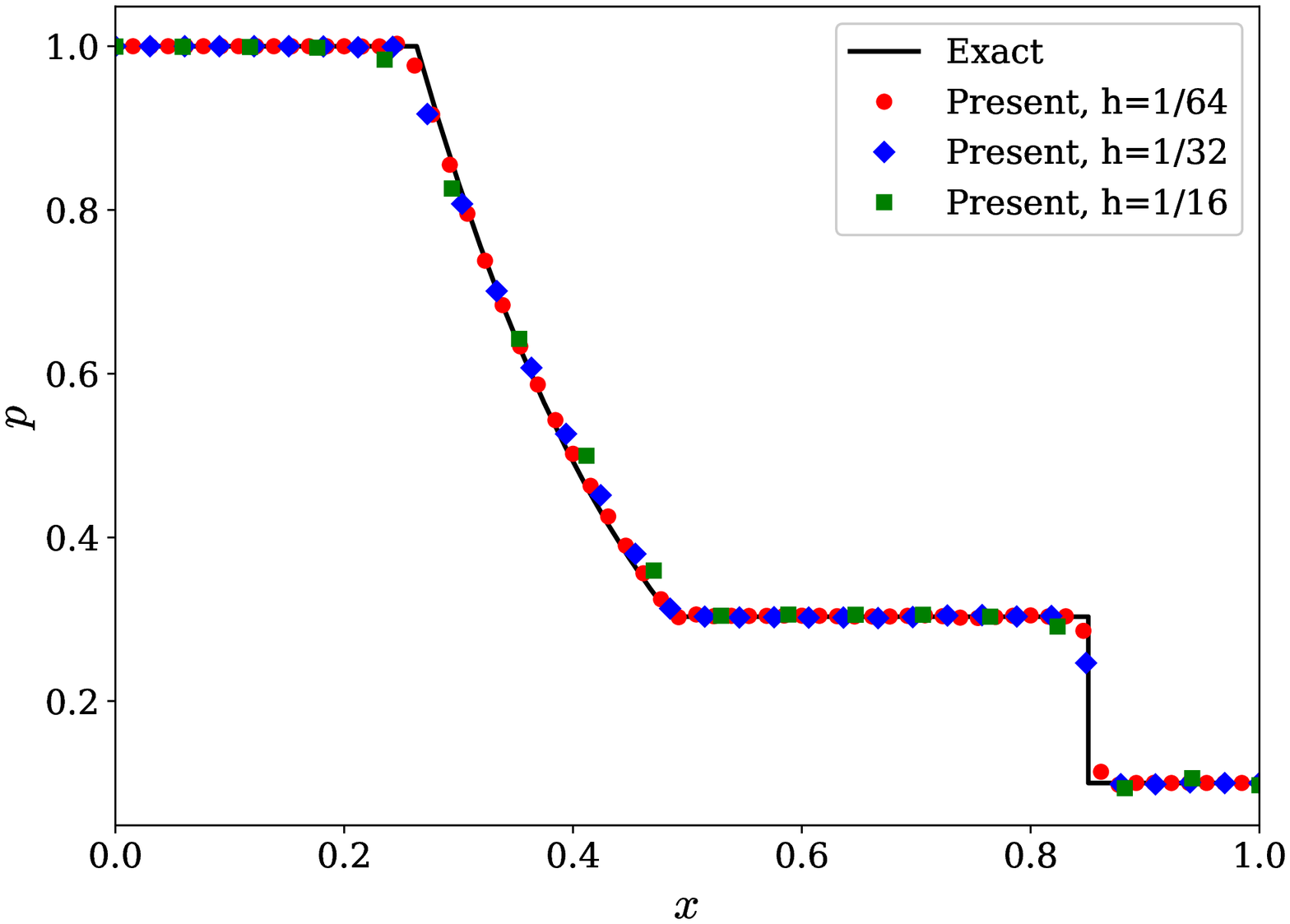}
		\label{fig:sod:p:h_refine}
	}
	\subfigure[]{
		\includegraphics[width=0.4 \textwidth]{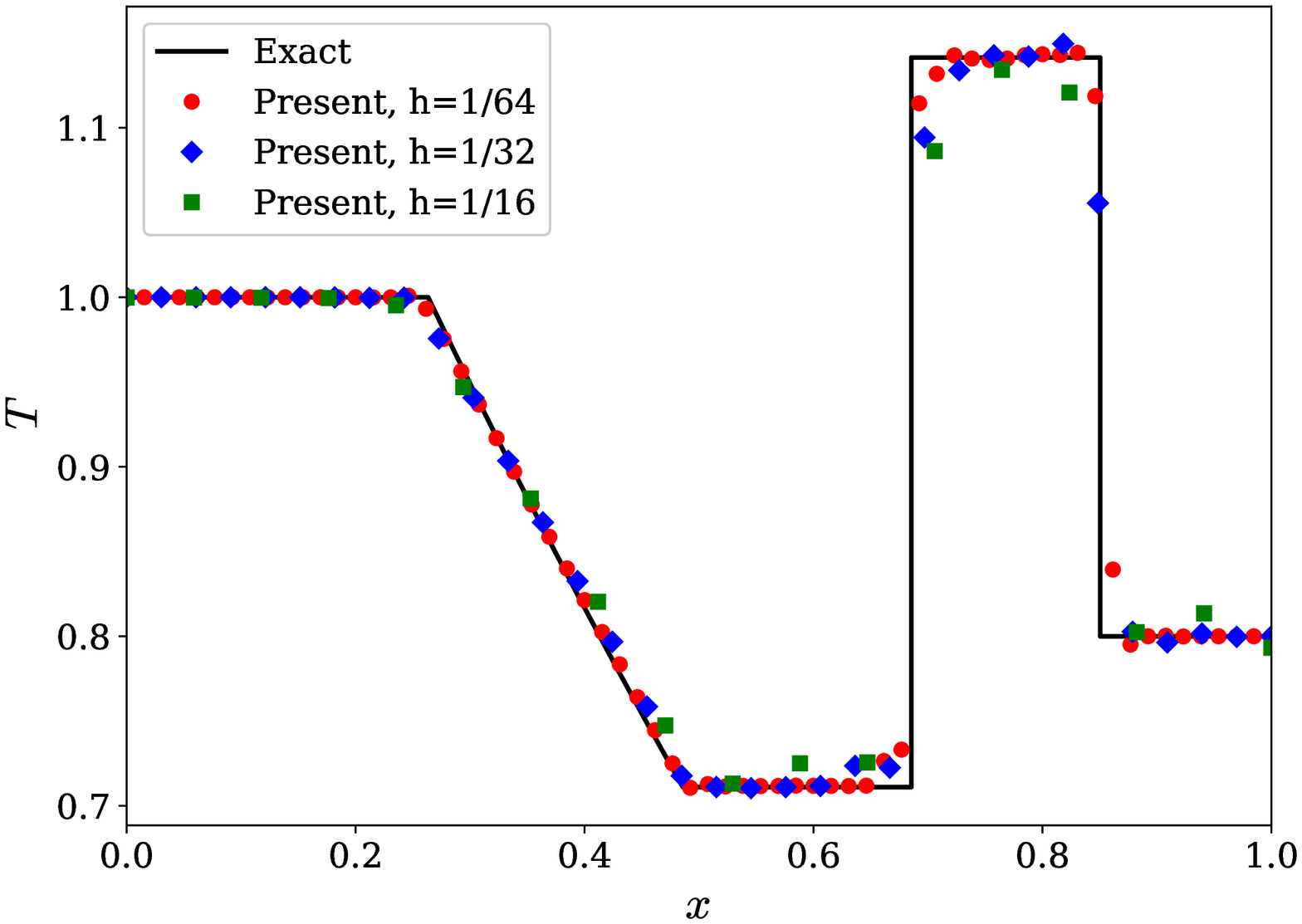}
		\label{fig:sod:T:h_refine}
	}
	\caption{The $h$ criterion grid refinement tests for Sod problem: (a) density, (b) u-velocity, (c) pressure,
		and (d) temperature distributions at $t = 0.2$.}
	\label{fig:sod:h_refine}
\end{figure}

\begin{figure}[!htp]
	\centering
	\subfigure[]{
		\includegraphics[width=0.4 \textwidth]{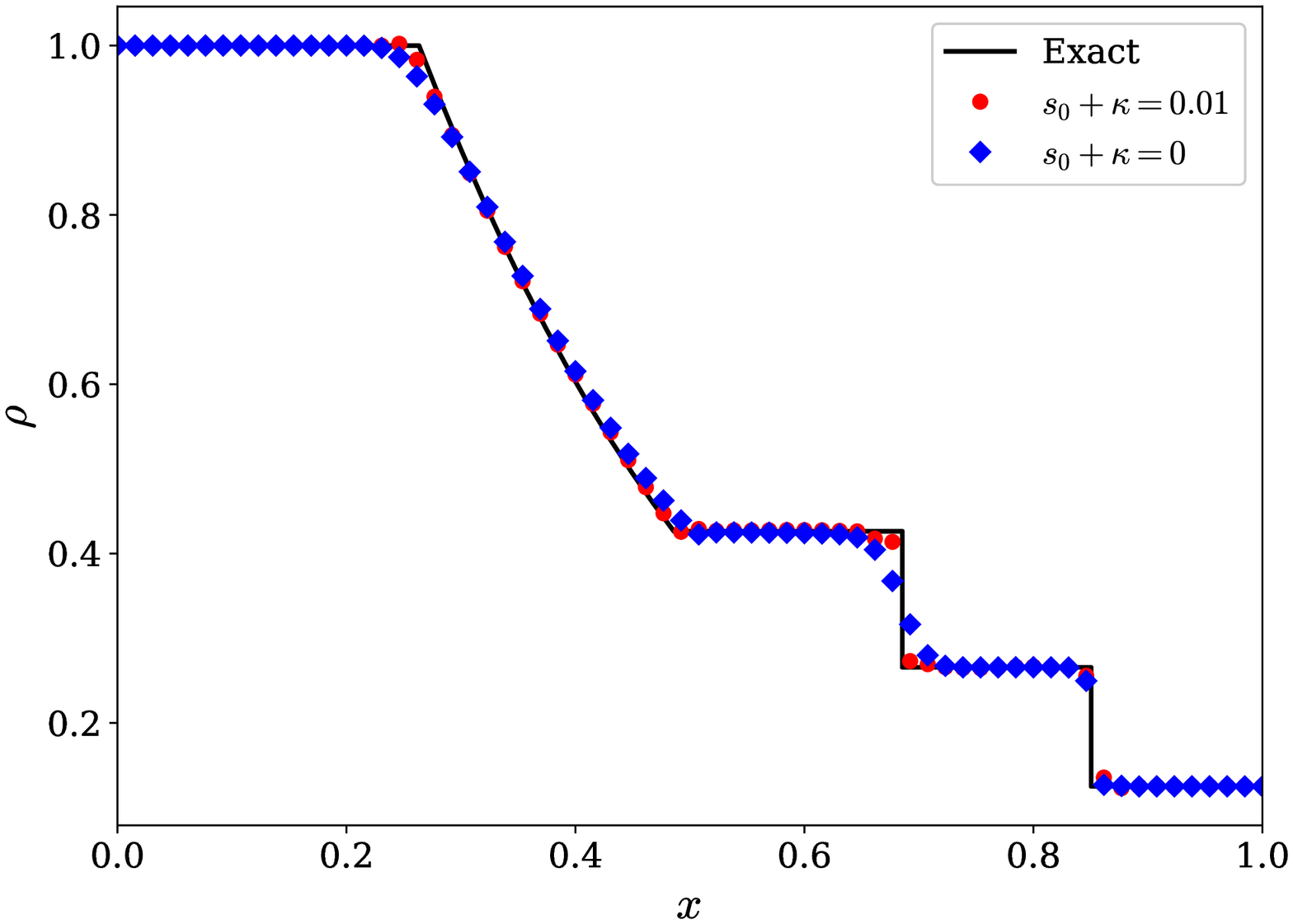}
		\label{fig:sod:rho:s0}
	}
	\subfigure[]{
		\includegraphics[width=0.4 \textwidth]{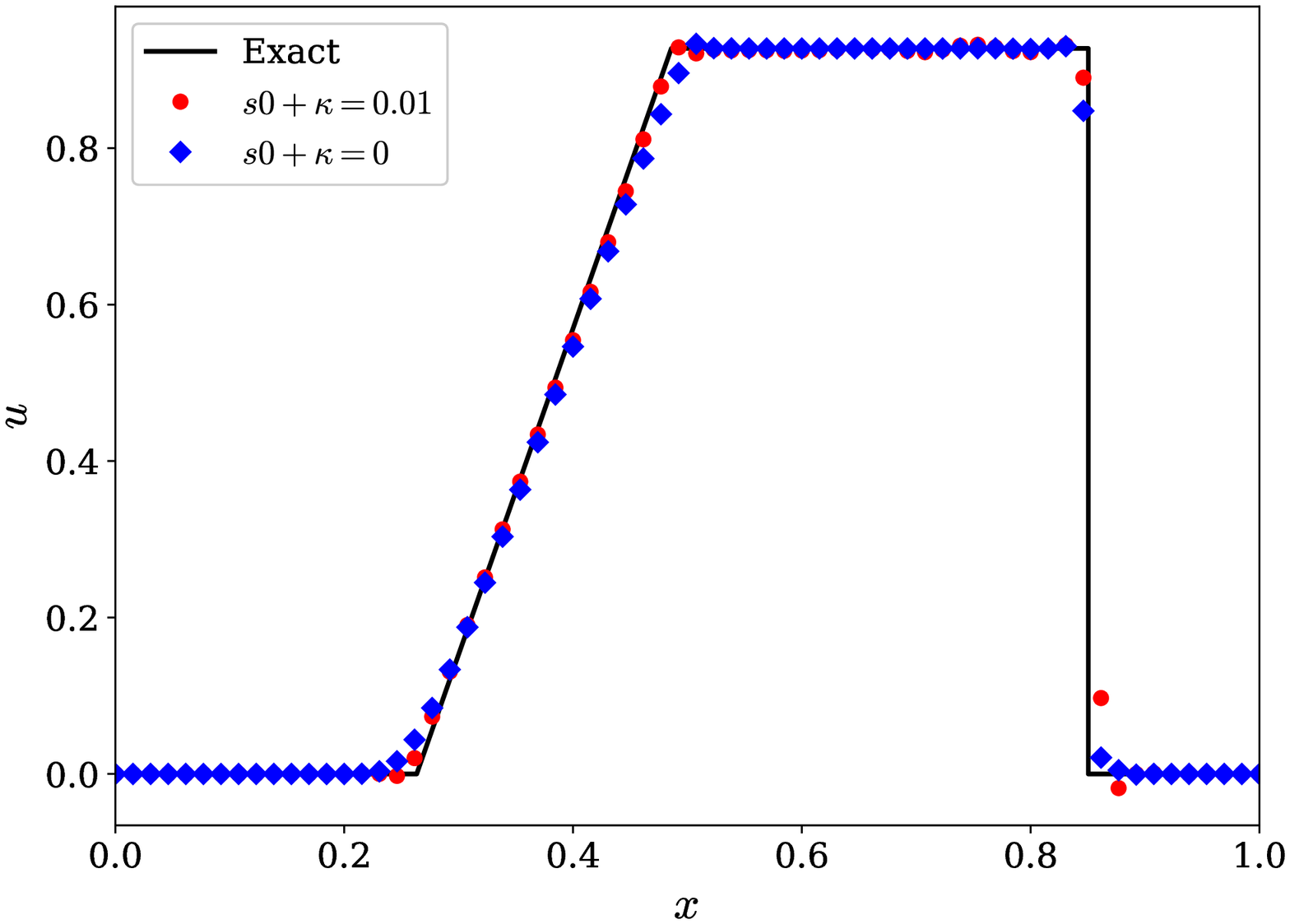}
		\label{fig:sod:u:s0}
	}
	\subfigure[]{
		\includegraphics[width=0.4 \textwidth]{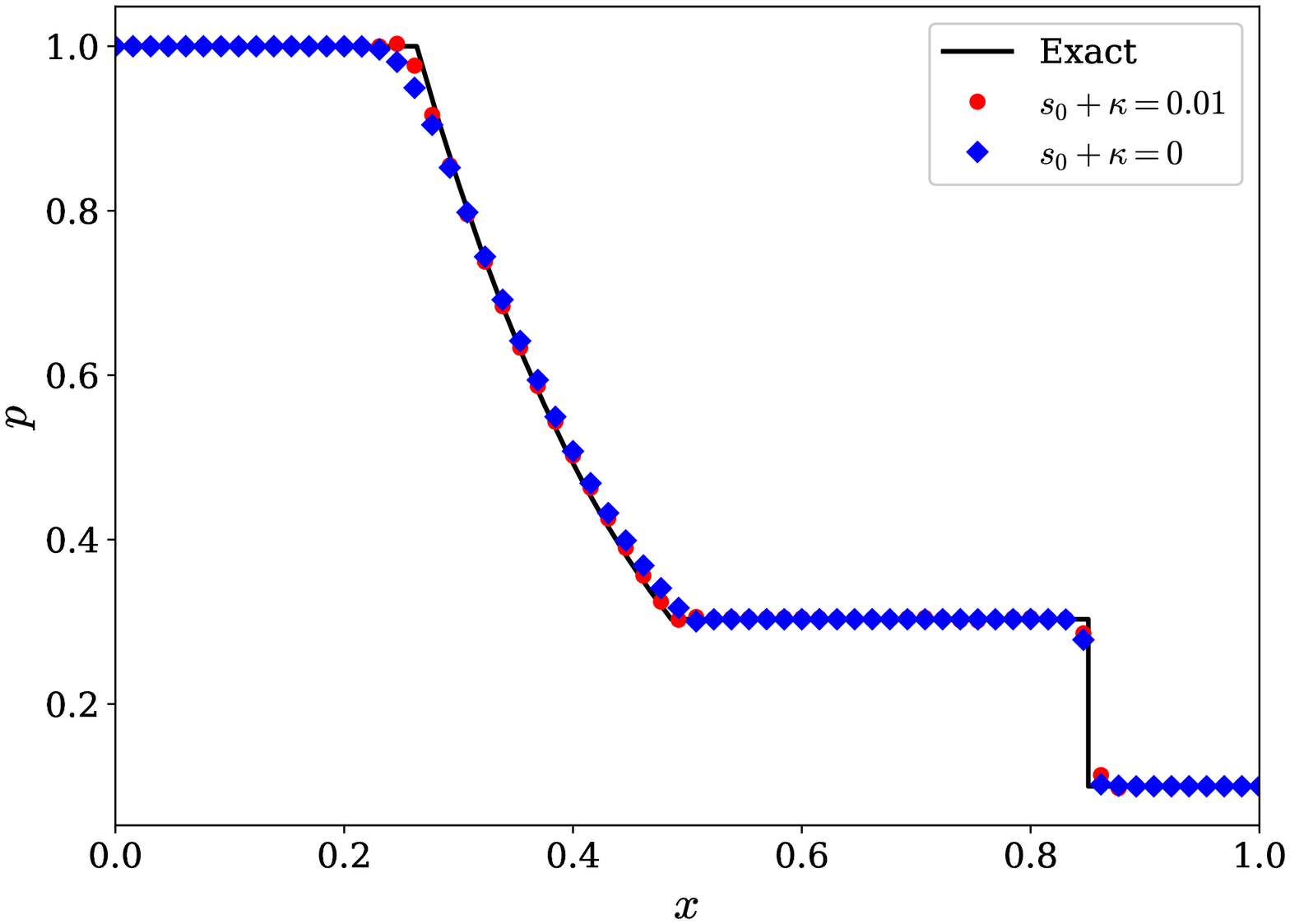}
		\label{fig:sod:p:s0}
	}
	\subfigure[]{
		\includegraphics[width=0.4 \textwidth]{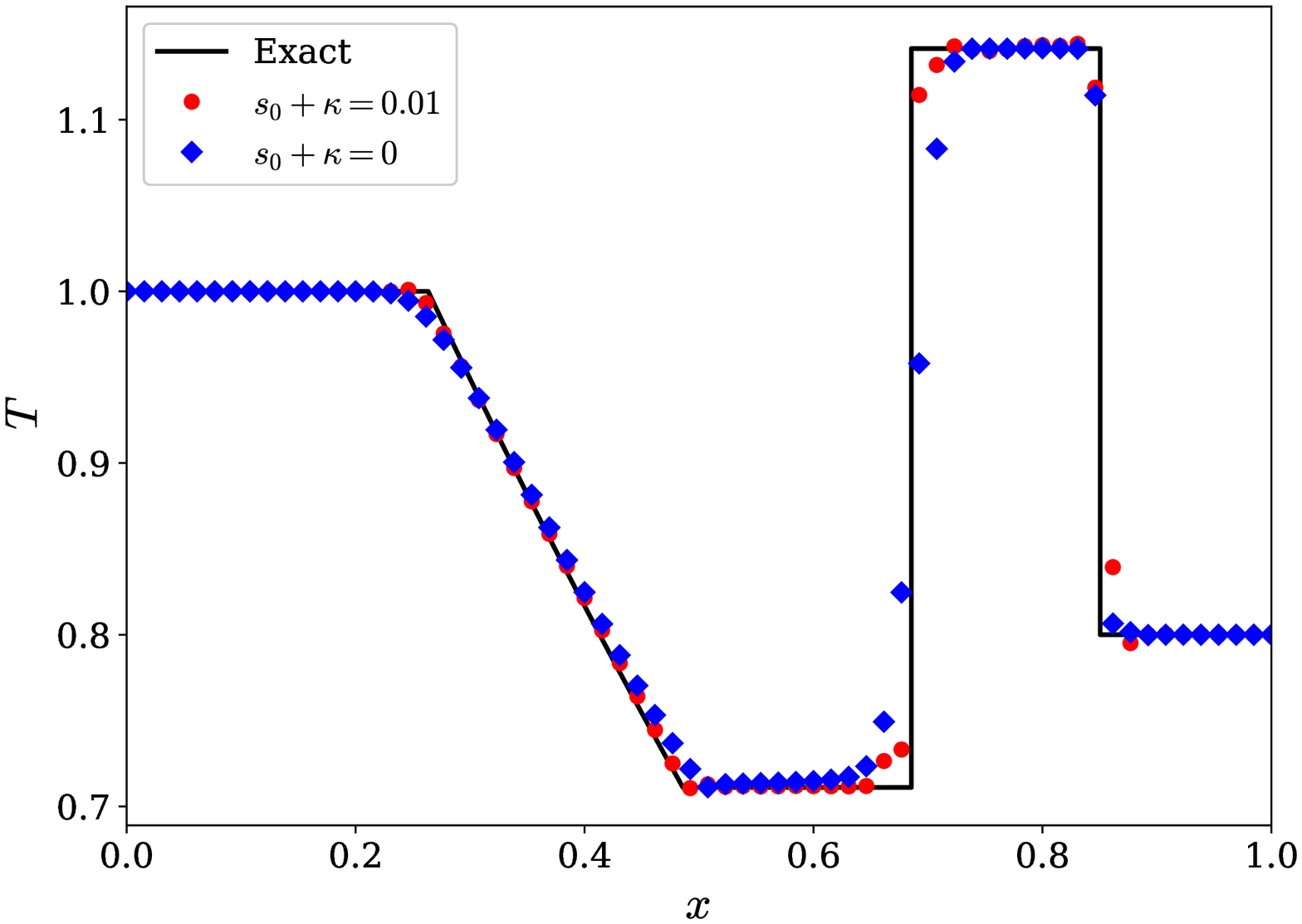}
		\label{fig:sod:T:s0}
	}
	\caption{The results of Sod problem with different values of $s_0+\kappa$: (a) density, (b) u-velocity, (c) pressure, and (d) temperature distributions at $t = 0.2$. The length scale $h$ is set as $1/64$.}
	\label{fig:sod:s0}
\end{figure}

The first one-dimensional Riemann problem is the Sod problem~\cite{SOD19781}, which is always used to validate the ability of numerical schemes to capture the discontinuity. The computational domain is $(x,y) \in [0, 1]\times[0,4h]$, and the Cartesian grid with different length scale $h$ is used in the approach. The initial condition reads
\begin{equation}\label{sodInit}
\left(\rho, u, v, p\right)= \left\{
\begin{array}{ll}
\left(1, 0, 0, 1\right),& 0<x<0.5,\\
\left(0.125, 0, 0, 0.1\right), & 0.5 \leq x \leq 1.
\end{array}
\right.
\end{equation}
Fig.~\ref{fig:sod:h_refine} shows the density, velocity, pressure, and temperature distributions at $t = 0.2$. The numerical results have a good accordance with the exact solution, and it can be obviously seen that the accuracy is improved with the $h$ criterion grid refinement.
For the shock capturing method used in present work, the value of $s_0+\kappa$ has a great effect to the numerical accuracy. It is known to all that the larger value of $s_0+\kappa$ is set, the less accuracy lost is obtained. Fig.~\ref{fig:sod:s0} plots the simulation with different values of $s_0+\kappa$. It is evident that the higher accuracy can be obtained with the larger value of $s_0+\kappa$.

The second one-dimensional Riemann problem is the Lax problem\cite{shu1988efficient}. Compared with Sod problem, Lax problem has a much stronger discontinuity. The computational domain is $(x,y) \in [0, 1]\times[0,4h]$, and the Cartesian grid with different length scale $h$ is used in the approach. The value of $s_0+\kappa$ is set as $0.01$ in the computation. \\
The initial condition is expressed as
\begin{equation}\label{initLax}
\left(\rho, u, v, p\right)= \left\{
\begin{array}{ll}
\left(0.445, 0.698, 0.0, 3.528\right), & 0<x<0.5,\\
\left(0.5, 0.0, 0.0, 0.571\right), & 0.5 \leq x \leq 1.
\end{array}
\right.
\end{equation}

\begin{figure}[!htp]
	\centering
	\subfigure[]{
		\includegraphics[width=0.4 \textwidth]{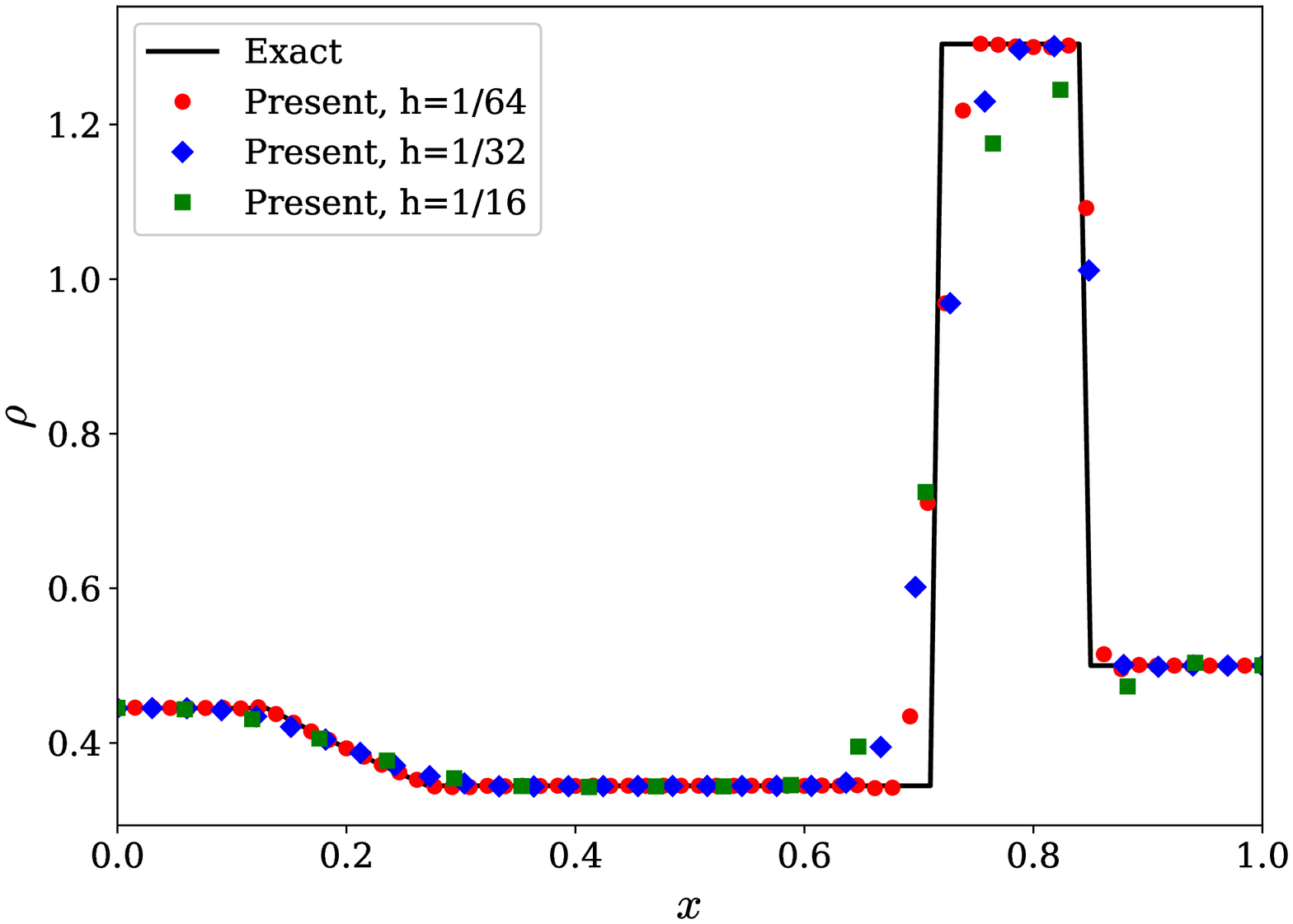}
		\label{fig:Lax:rho:h_refine}
	}
	\subfigure[]{
		\includegraphics[width=0.4 \textwidth]{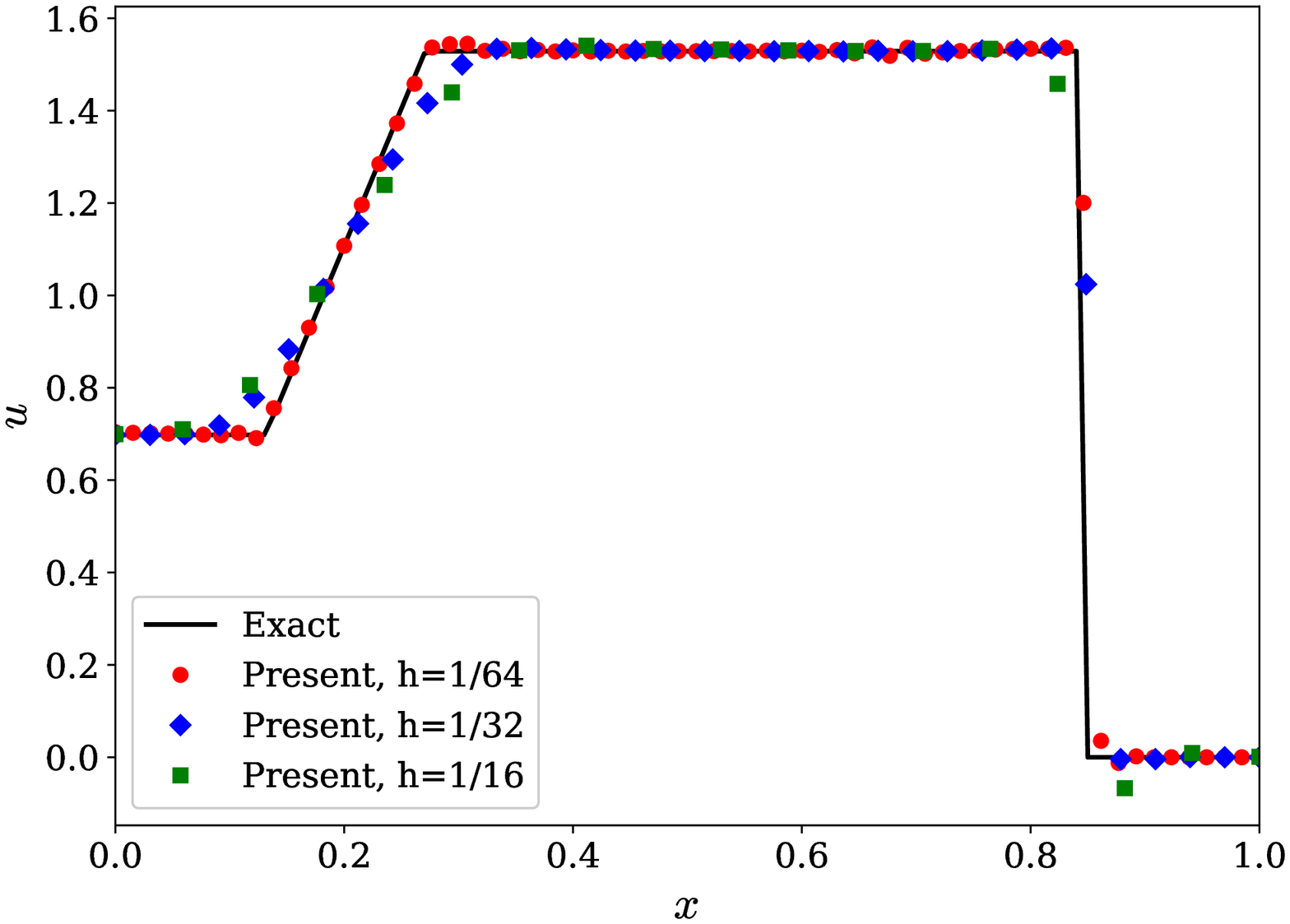}
		\label{fig:Lax:u:h_refine}
	}
	\subfigure[]{
		\includegraphics[width=0.4 \textwidth]{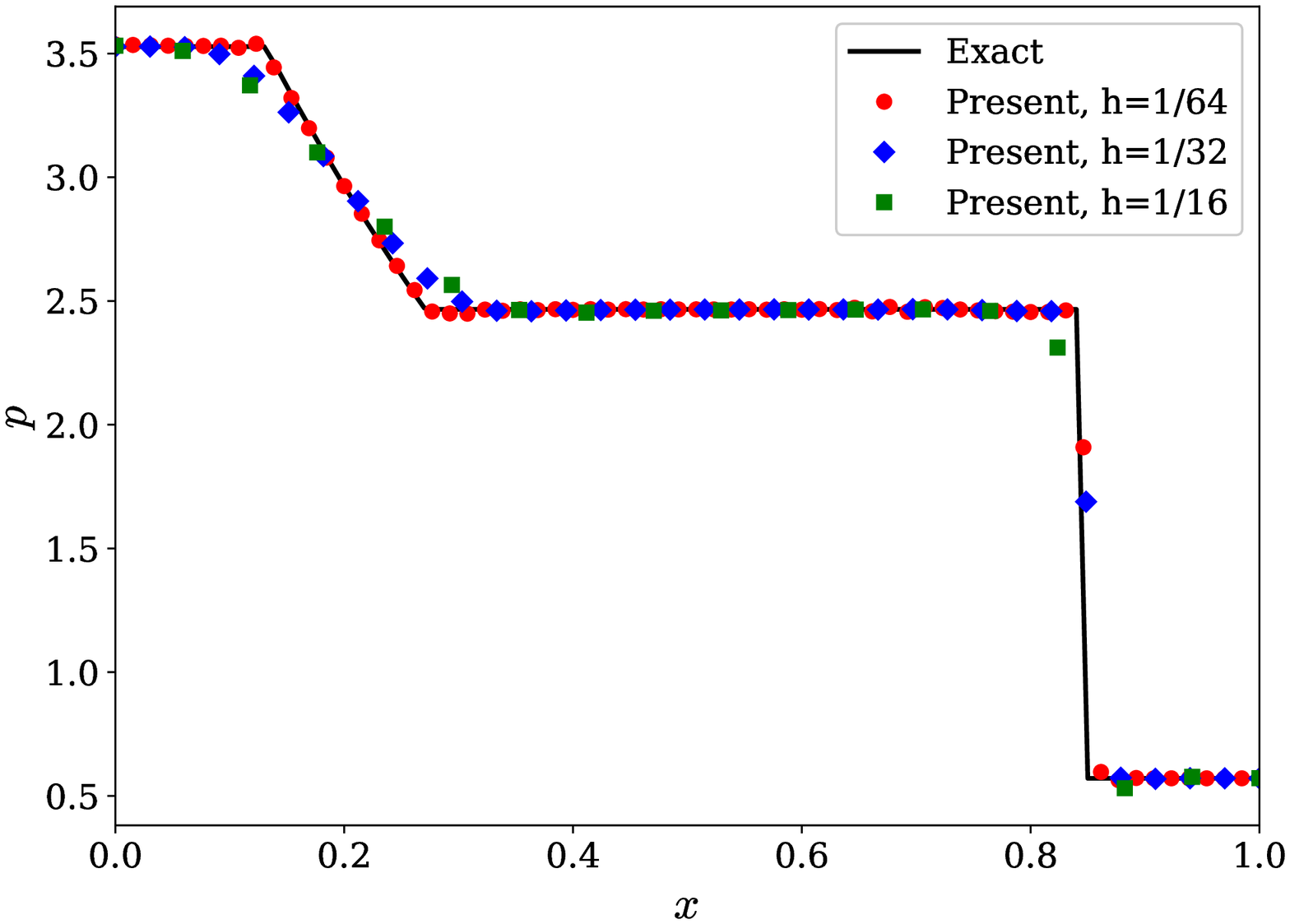}
		\label{fig:Lax:p:h_refine}
	}
	\subfigure[]{
		\includegraphics[width=0.4 \textwidth]{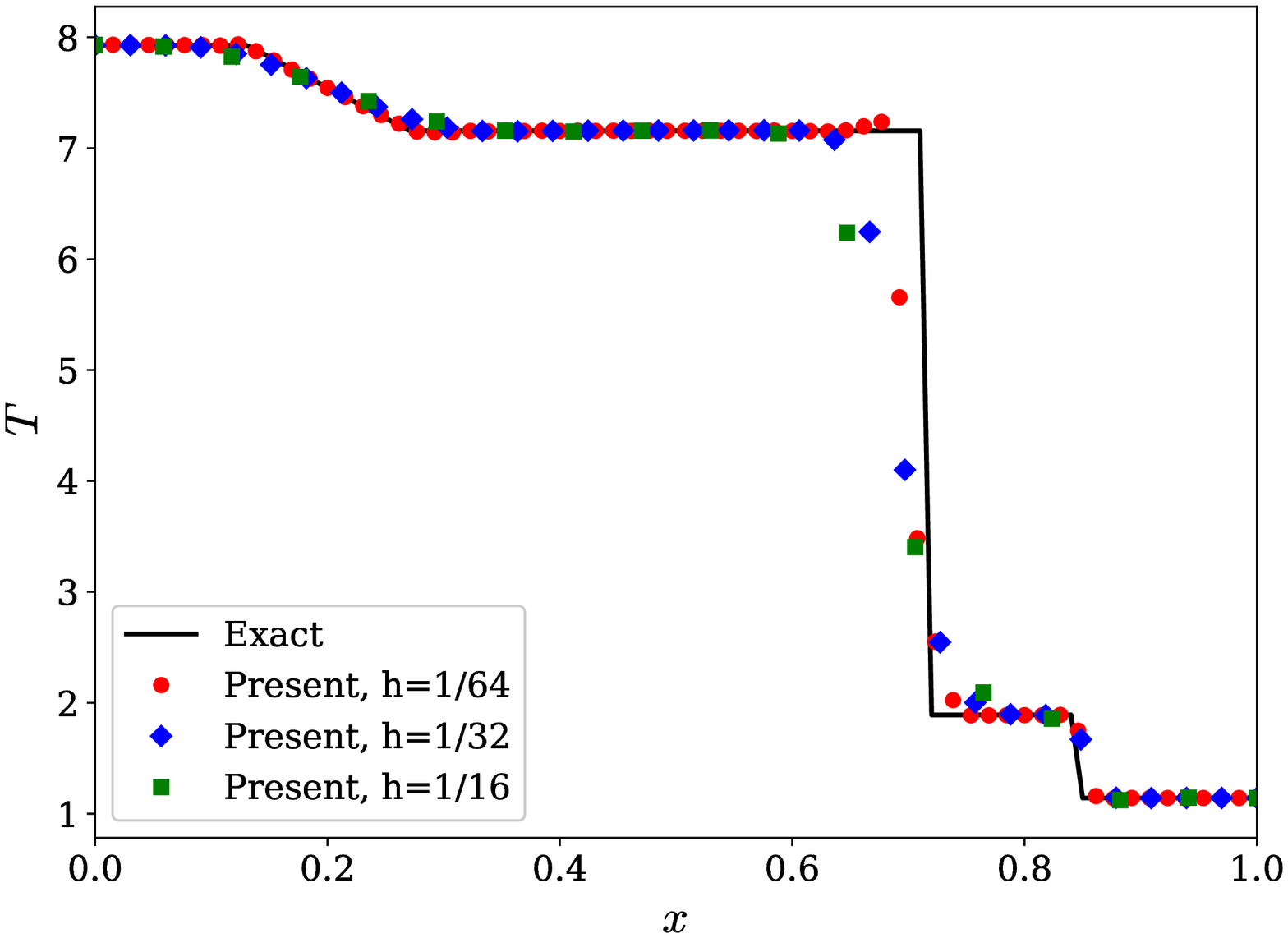}
		\label{fig:Lax:T:h_refine}
	}
	\caption{The $h$ criterion grid refinement tests for Lax problem: (a) density, (b) u-velocity, (c) pressure, 
		and (d) temperature distributions at $t=0.14$.}
	\label{fig:Lax:h_refine}
\end{figure}

Fig.~\ref{fig:Lax:h_refine} shows the density, velocity, pressure and temperature distributions at $t = 0.14$. The numerical results have a good accordance with the exact solution. The little oscillation can be seen nearby the discontinuity, and it is because that the shock is not captured very well. The shock capturing method is an open question needed to be further studied.

\subsection{Shu-Osher problem}
The problem of Shu-Osher~\cite{shu1988efficient} describes the interaction of a sinusoidal density-wave with a Mach 3 normal shock. The purpose of this case is to validate the behavior of our method on the shock-wave interaction problem. The shock capturing method is also examined in the case, and $s_0+\kappa$ is set as $0.01$ in the simulation. The computational domain used in the simulation is taken as $[0, 10]\times[0,4h]$. The Cartesian grid is used in the simulations, and the length scale of the grid is $h=1/100$. The upper and the bottom boundaries are set as the periodic boundary, and the left and right side are set as the non-reflecting boundary. The initial condition is given as
\begin{equation}\label{initShuOsher}
	\left(\rho, u, v, p\right)=\left\{
	\begin{array}{ll}
	\left(3.857143, 2.629369, 0.0, 10.33333\right), & 0 \leq x < 1,\\
	\left(1+0.2sin(5(x-4)), 0.0, 0.0, 1.0\right), & 1 \leq x \leq 10.
	\end{array}
\right.
\end{equation}
 \begin{figure}
	\centering
	\includegraphics[width=0.9 \textwidth]{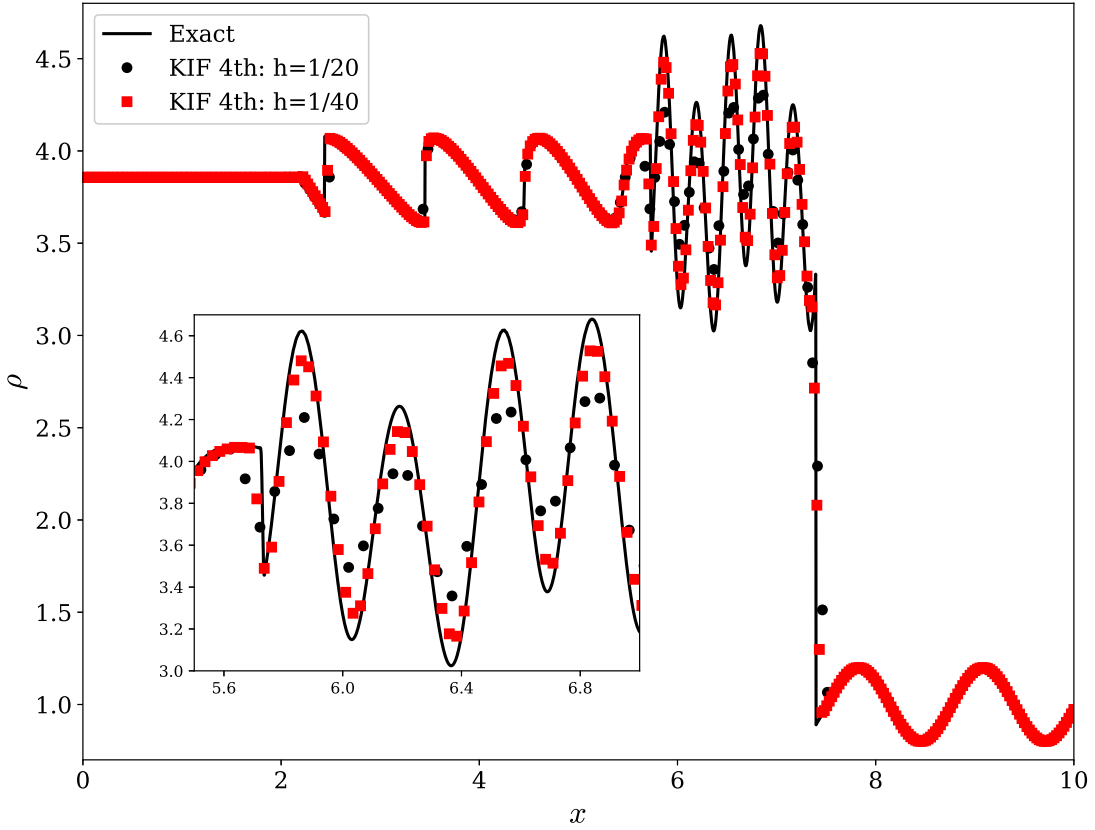}
	\caption{\label{fig:shu-osher} Shu-Osher problem: the distribution of density-wave at $t = 1.8$.}
\end{figure}
Fig.~\ref{fig:shu-osher} shows the density distribution at $t = 1.8$, and the zoom in view near the high frequency wave is also exhibited. Because the exact solution of this problem can not be computed directly, the solution of fourth order WENO method with $10000$ grid points in one dimension is taken as the exact result. It can be seen in Fig.~\ref{fig:shu-osher} that the performance is improved with increasing mesh resolution. The discontinuity is captured well, and the ability of present scheme to capture the frequency wave is also be verified in the case.

\subsection{Shock vortex interaction problem}
The shock vortex interaction problem~\cite{shu1998essentially} is always used to validate the performance of high order method. Compared to lower order method, the high order scheme have the advantages on resolving the vortex and interaction.\\
In the simulation, a stationary normal shock and a small perturbation are initialed in the flow field. A Mach 1.1 normal shock wave is located at the position $x=0.5$. The left side state ($Ma=1.1$) of the shock wave is given as follows
\begin{equation}\label{shockVortexLeft}
(\rho,u,v,p)=(Ma^2, \sqrt{\gamma}, 0.0, 1.0),\quad T=p/\rho,\quad S=ln(p/\rho^{\gamma}).
\end{equation}
A small and weak vortex is superposed to the left side of the normal shock. The center of the vortex is $(x_c, y_c)=(0.25, 0.5)$. The perturbation is given as
\begin{equation}\label{shockVortexPerturbation1}
(\delta u, \delta v) = \kappa\eta e^{\mu(1-\eta^2)}(sin\theta,-cos\theta),
\end{equation}
\begin{equation}\label{shockVortexPerturbation2}
\delta T = -\frac{(\gamma-1)\kappa^2}{4\mu\gamma}e^{2\mu(1-\eta^2)}(sin\theta,-cos\theta), \delta S=0,
\end{equation}
where
\begin{equation}\label{shockvortexcoefficients}
\begin{aligned}
\kappa &= 0.3, \\
\mu &= 0.204, \\
\eta &= r/r_c, \\
r_c &= 0.05, \\
r &= \sqrt{(x-x_c)^2+(y-y_c)^2}.
\end{aligned}
\end{equation}

\begin{figure}[!htp]
	\includegraphics[width=0.4 \textwidth]{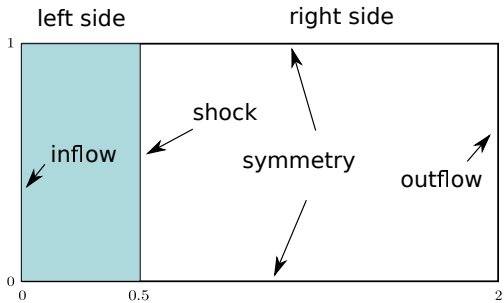}
	\caption{\label{fig:Shock_Vortex:domain} Shock vortex interaction problem: the computational domain and the boundary conditions. In the simulation, Cartesian grid is used, and the element length scale of mesh $h$ equals to $1/100$.}
\end{figure}

\begin{figure}[!htp]
	\centering
	\subfigure[]{
		\includegraphics[width=0.4 \textwidth]{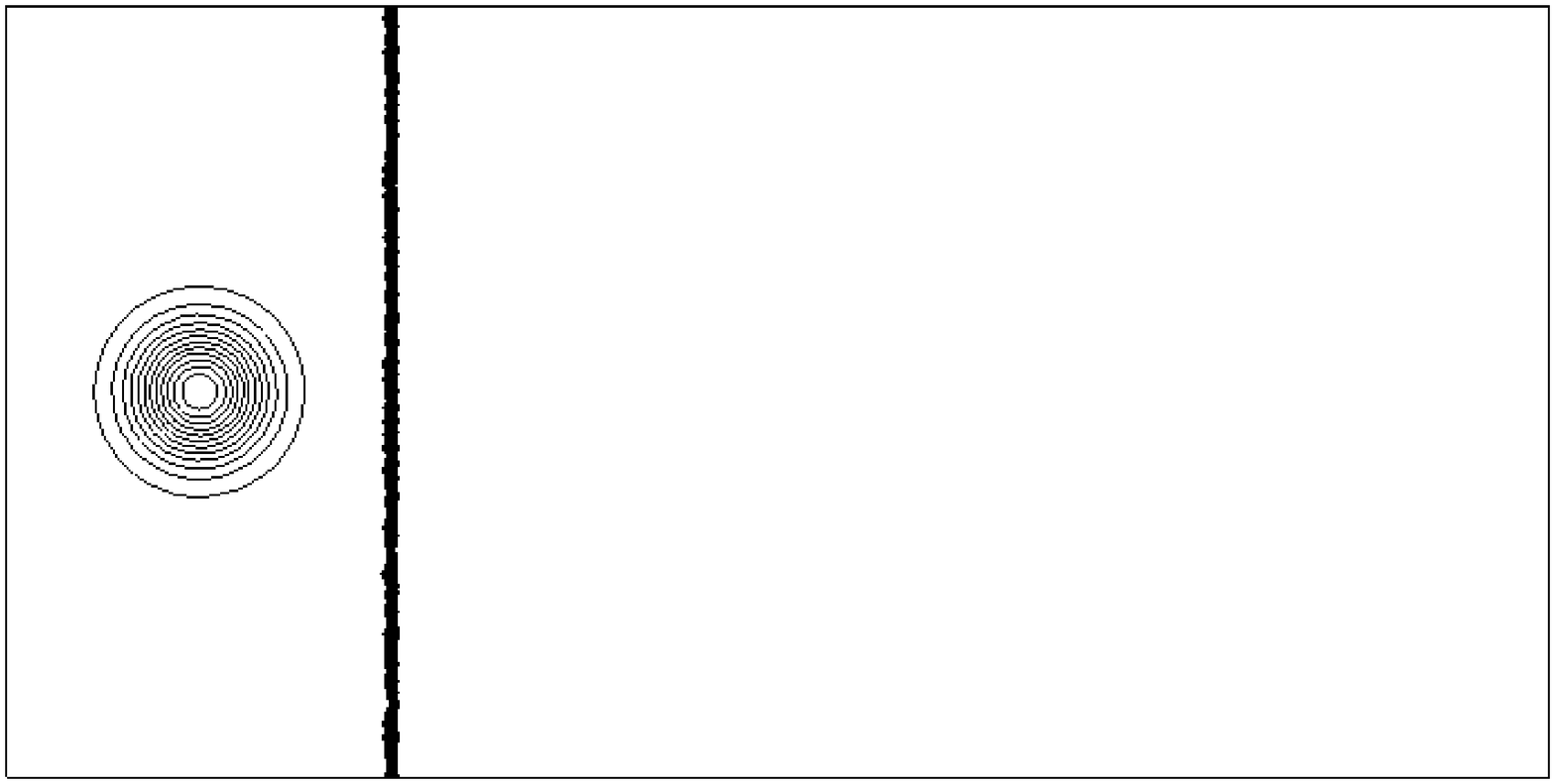}
		\label{fig:Shock_Vortex:t1}
	}
	\subfigure[]{
		\includegraphics[width=0.4 \textwidth]{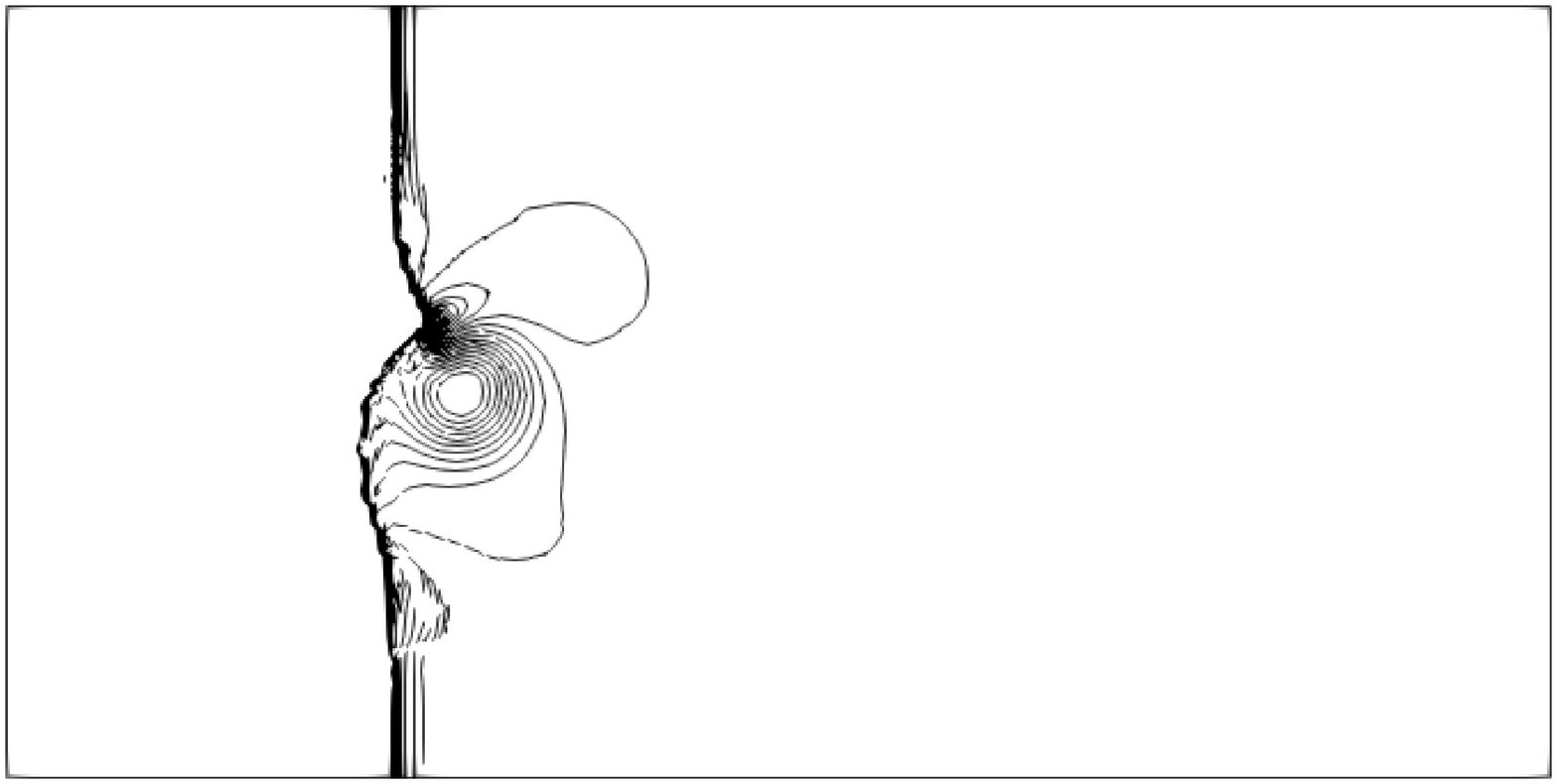}
		\label{fig:Shock_Vortex:t2}
	}
	\caption{\label{fig:Shock_Vortex:t1t2} Shock vortex interaction problem: (a) the initial status and (b) the contour of pressure at $t = 0.3$.}
\end{figure}

\begin{figure}[!htp]
	\centering
	\subfigure[]{
		\includegraphics[width=0.4 \textwidth]{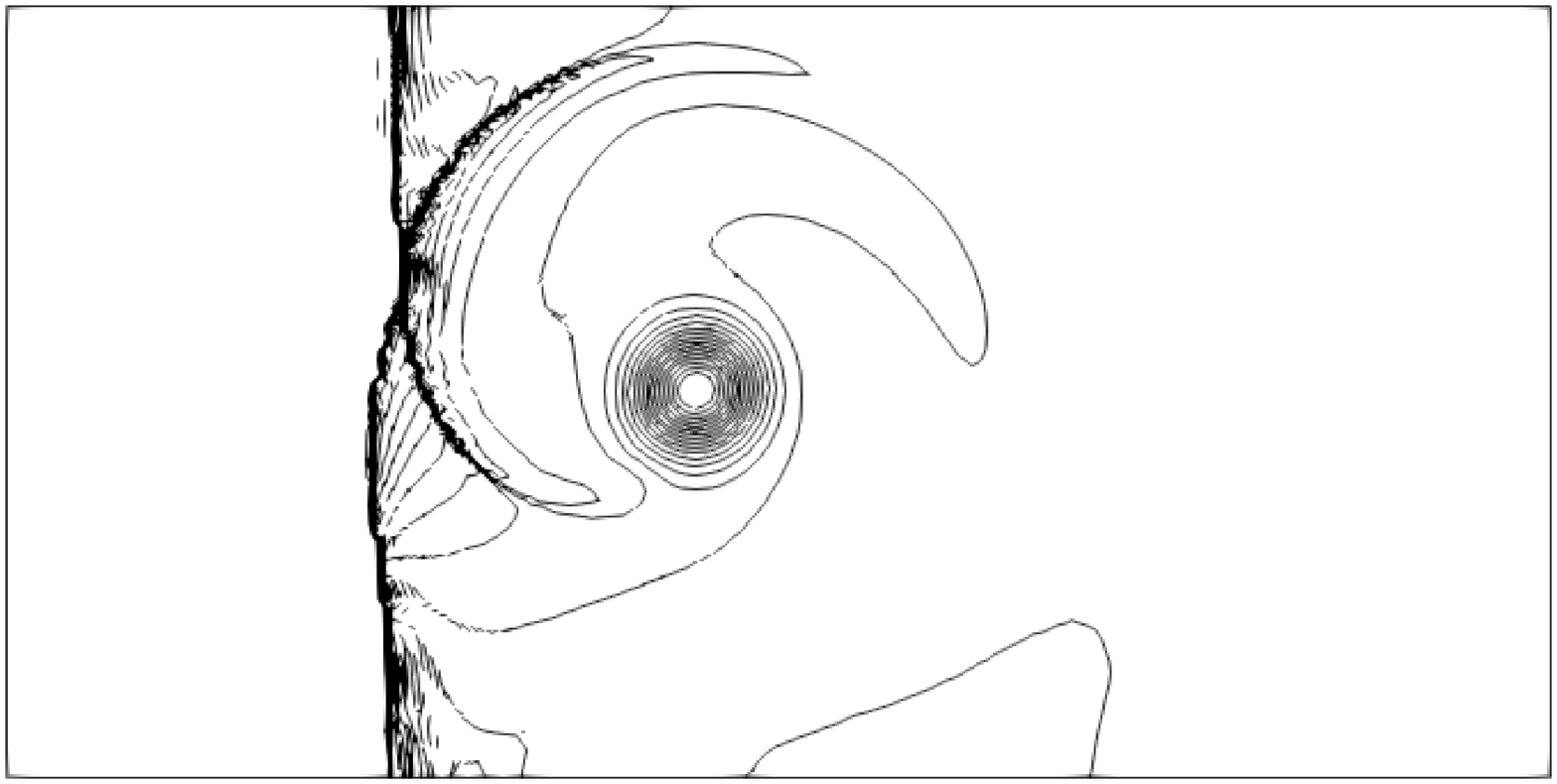}
		\label{fig:Shock_Vortex:t3}
	}
	\subfigure[]{
		\includegraphics[width=0.4 \textwidth]{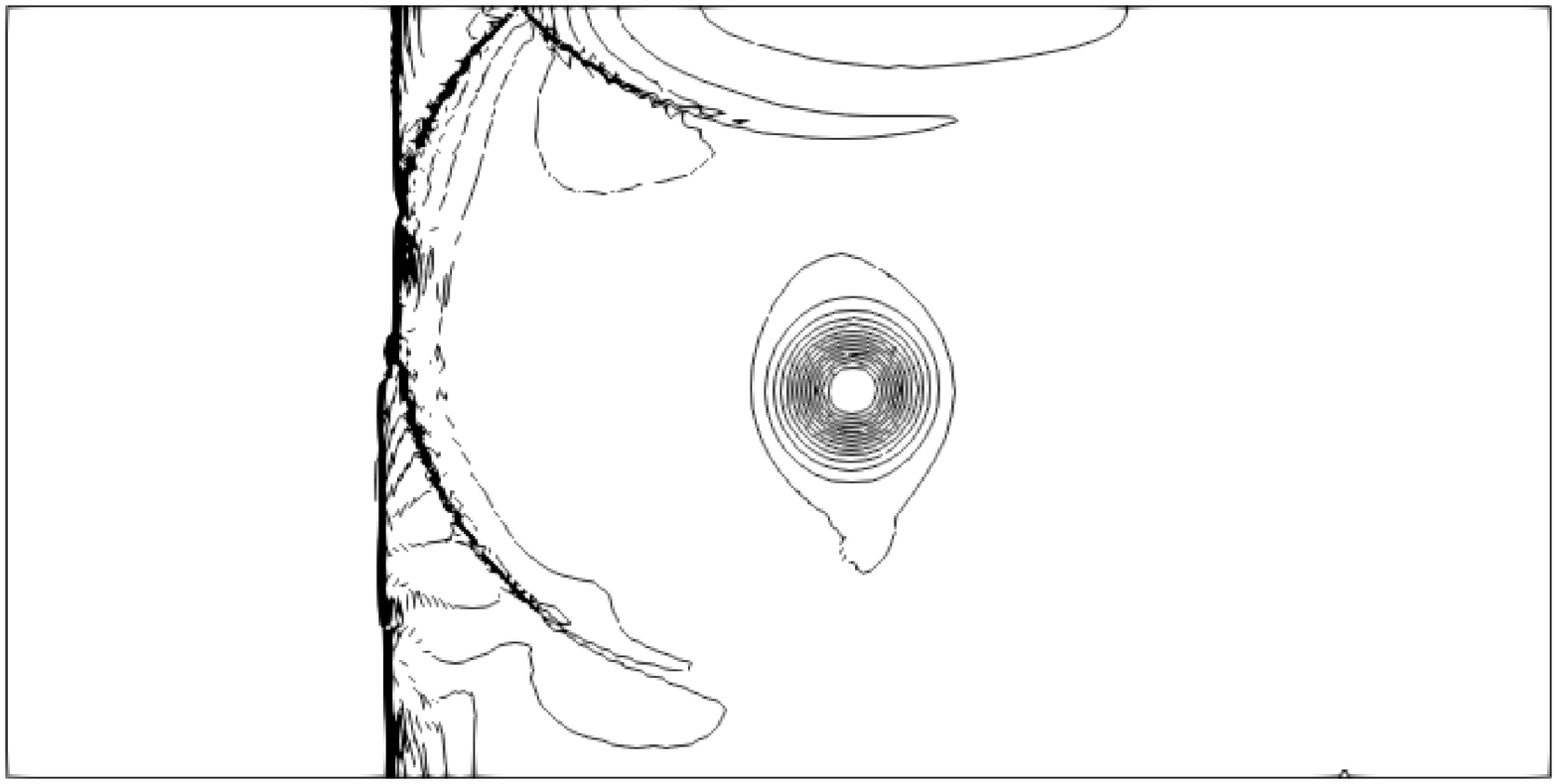}
		\label{fig:Shock_Vortex:t4}
	}
	\caption{\label{fig:Shock_Vortex:t3t4} Shock vortex interaction problem: (a) the contours of pressure at $t = 0.6$
		and (b) the contours of pressure at $t = 0.8$.}
\end{figure}

The computational domain and boundary conditions are exhibited in the Fig.~\ref{fig:Shock_Vortex:domain}. In the simulation, the Cartesian grid is used and the grid size $h$ is $1/100$. To capture the discontinuity, the coefficient $s_0+\kappa$ is set as $0.01$. The initial status at $t = 0$ and the contour plots at $t = 0.3$, $t = 0.6$ and $t = 0.8$ are shown in Fig.~\ref{fig:Shock_Vortex:t1t2} and Fig.~\ref{fig:Shock_Vortex:t3t4}. The perturbation is initialed
at $t=0$, and then the vortex moves from left to right across the shock. The profile of vortex varies with the movement, and the interaction of shock and vortex can be seen obviously in the Fig.~\ref{fig:Shock_Vortex:t1t2} and Fig.~\ref{fig:Shock_Vortex:t3t4}. The plots show that our present method can capture the shock and vortex interaction with enough resolution, and the vortex is recovered well. It also shows clearly in Fig.~\ref{fig:Shock_Vortex:t3t4} that the shock bifurcations reaches to the top boundary, and the reflection is evident.

\subsection{Lid-driven cavity flow}
\begin{figure}
	\centering
	\includegraphics[width=0.4 \textwidth]{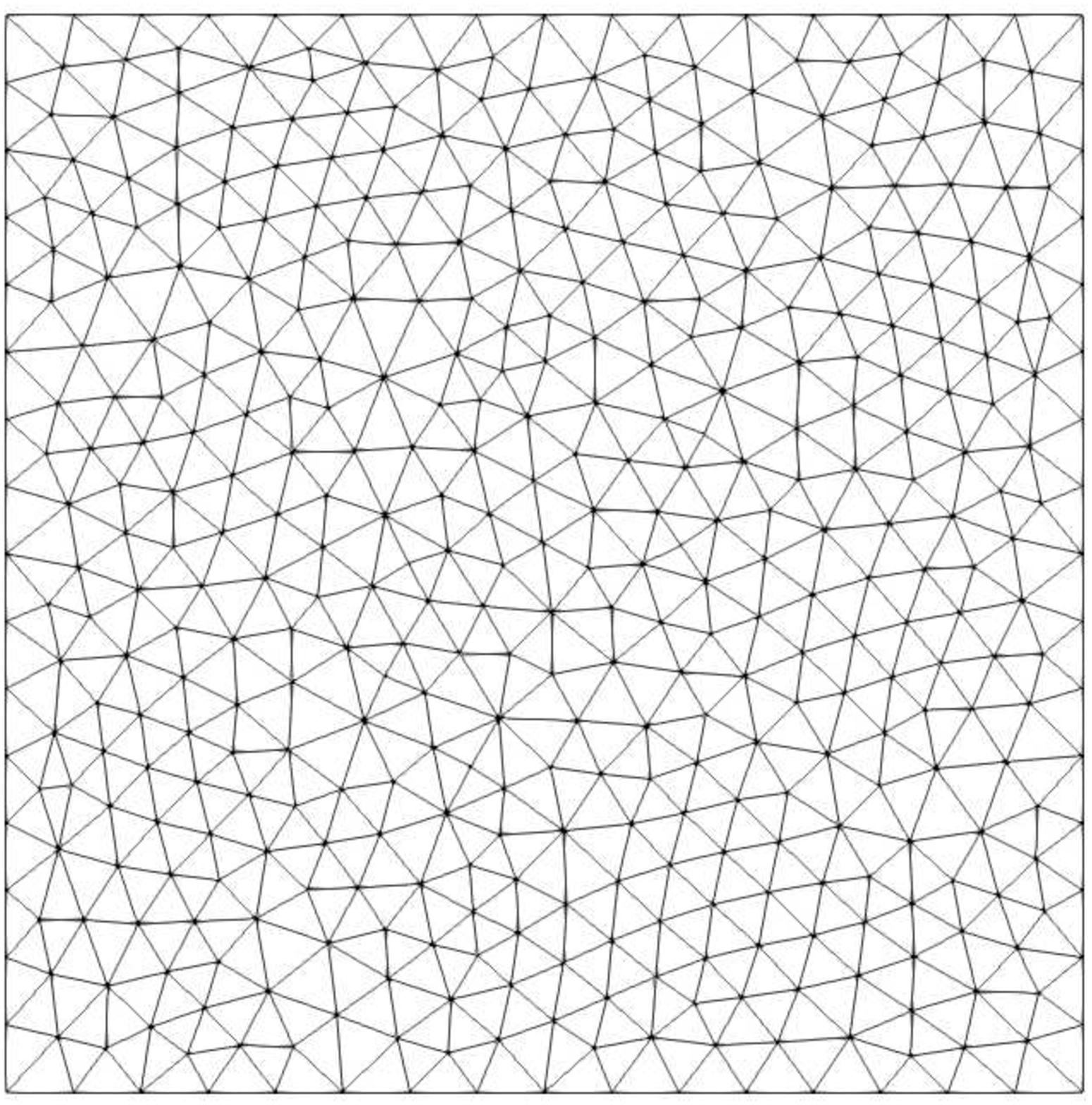}
	\caption{\label{fig:cavity:grid_tri} The triangular grid used in the simulation of lid-driven cavity flow (816 triangular elements).}
\end{figure}

\begin{figure}[!htp]
	\centering
	\subfigure[]{
		\includegraphics[width=0.4 \textwidth]{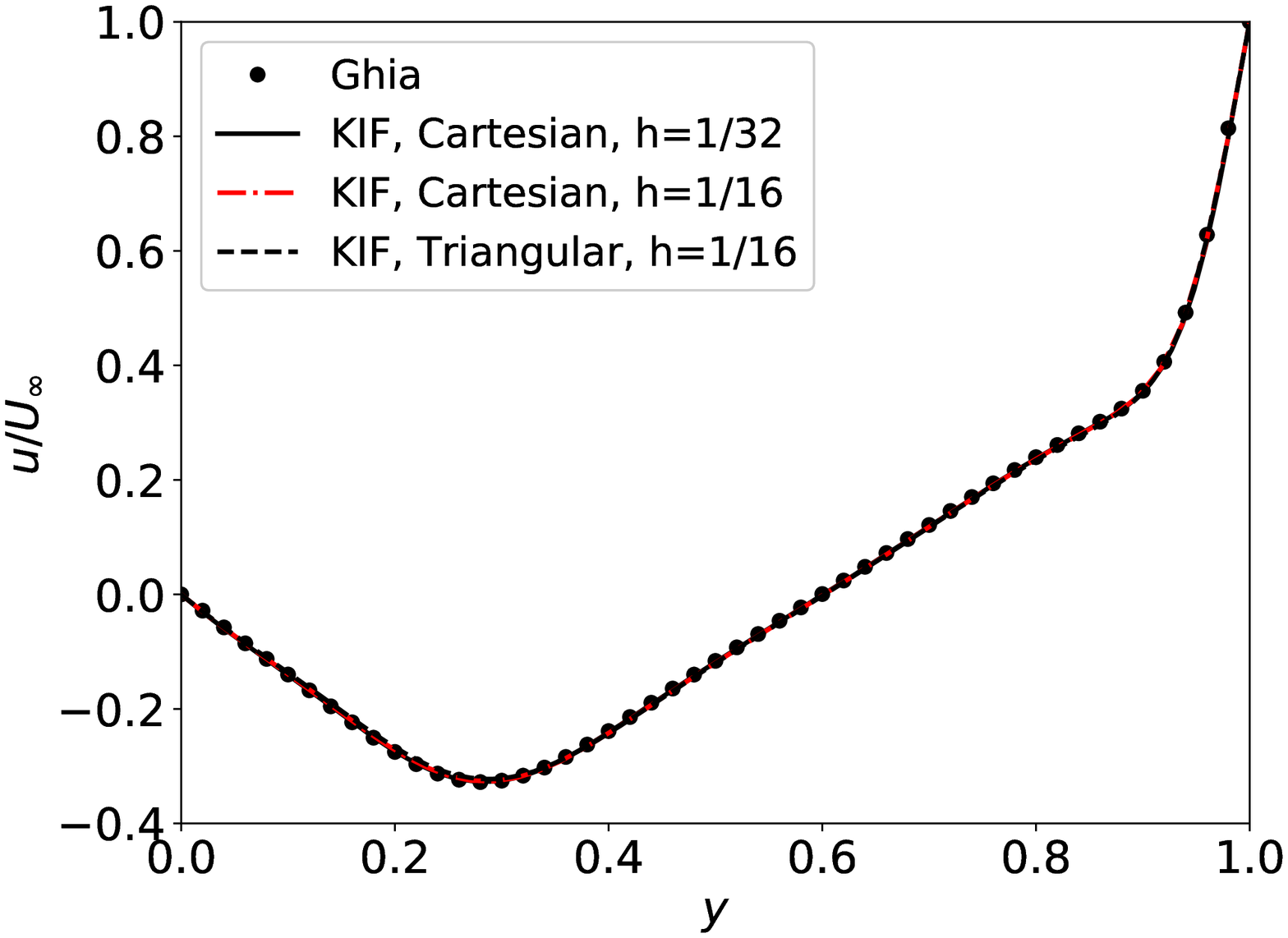}
		\label{fig:cavity:u:re400}
	}
	\subfigure[]{
		\includegraphics[width=0.4 \textwidth]{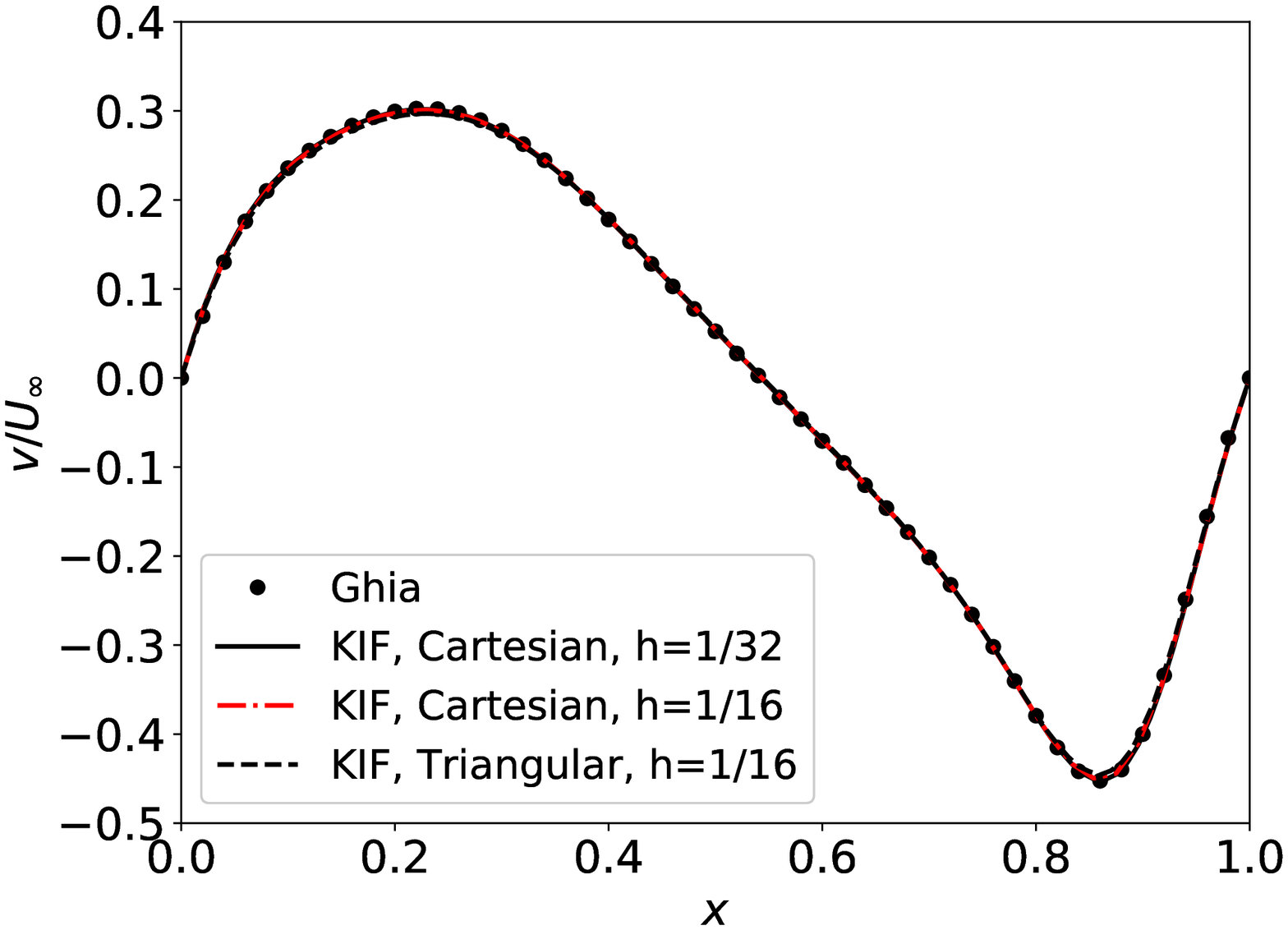}
		\label{fig:cavity:v:re400}
	}
	\caption{The velocity profile of the lid driven cavity flow at $Re = 400$: (a) u-velocity profiles at $x = 0.5$ and (b) v-velocity profiles at $y = 0.5$.}
	\label{fig:cavity:uv:re400}
\end{figure}

\begin{figure}[!htp]
	\centering
	\subfigure[]{
		\includegraphics[width=0.4 \textwidth]{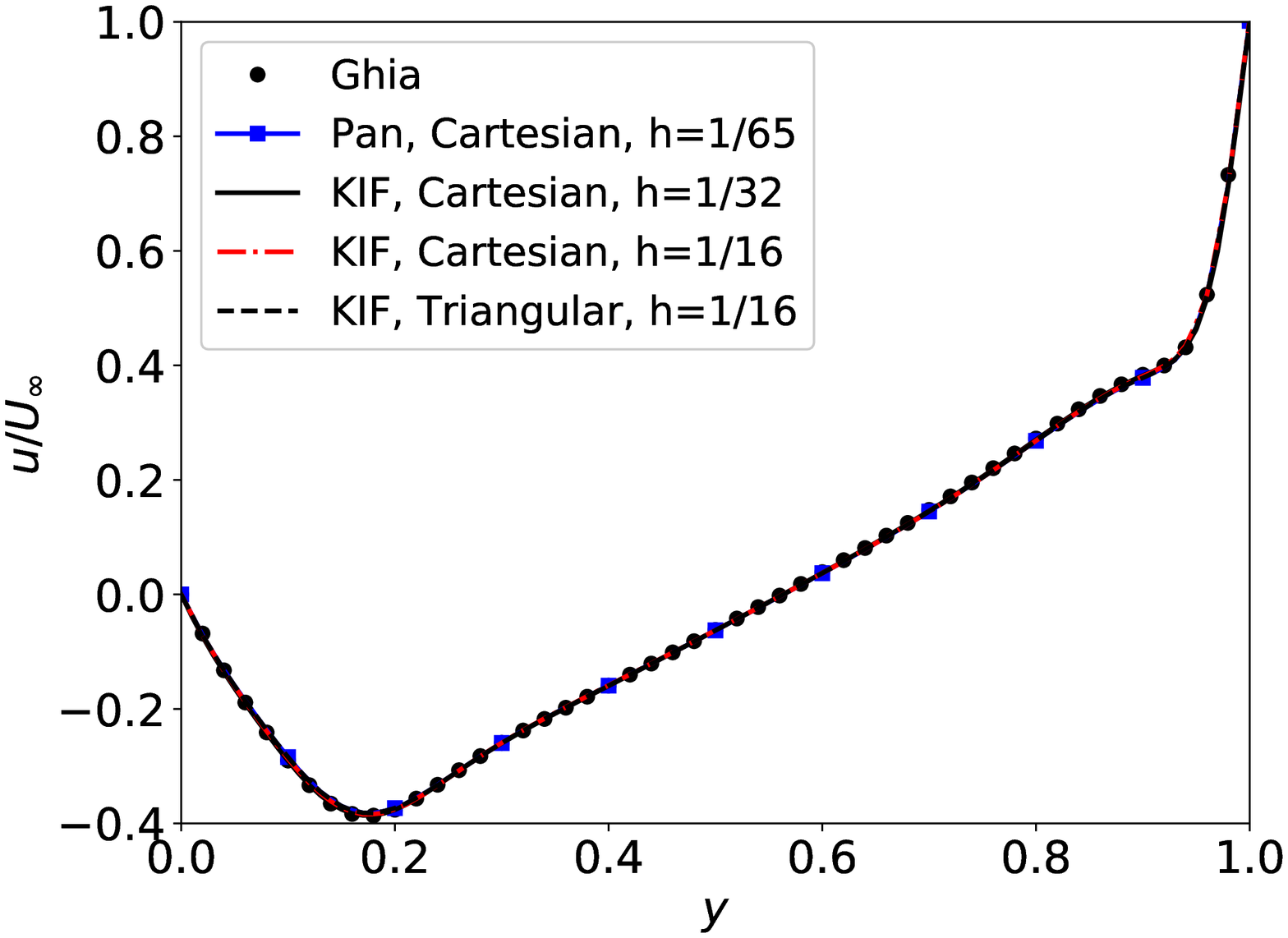}
		\label{fig:cavity:u:re1000}
	}
	\subfigure[]{
		\includegraphics[width=0.4 \textwidth]{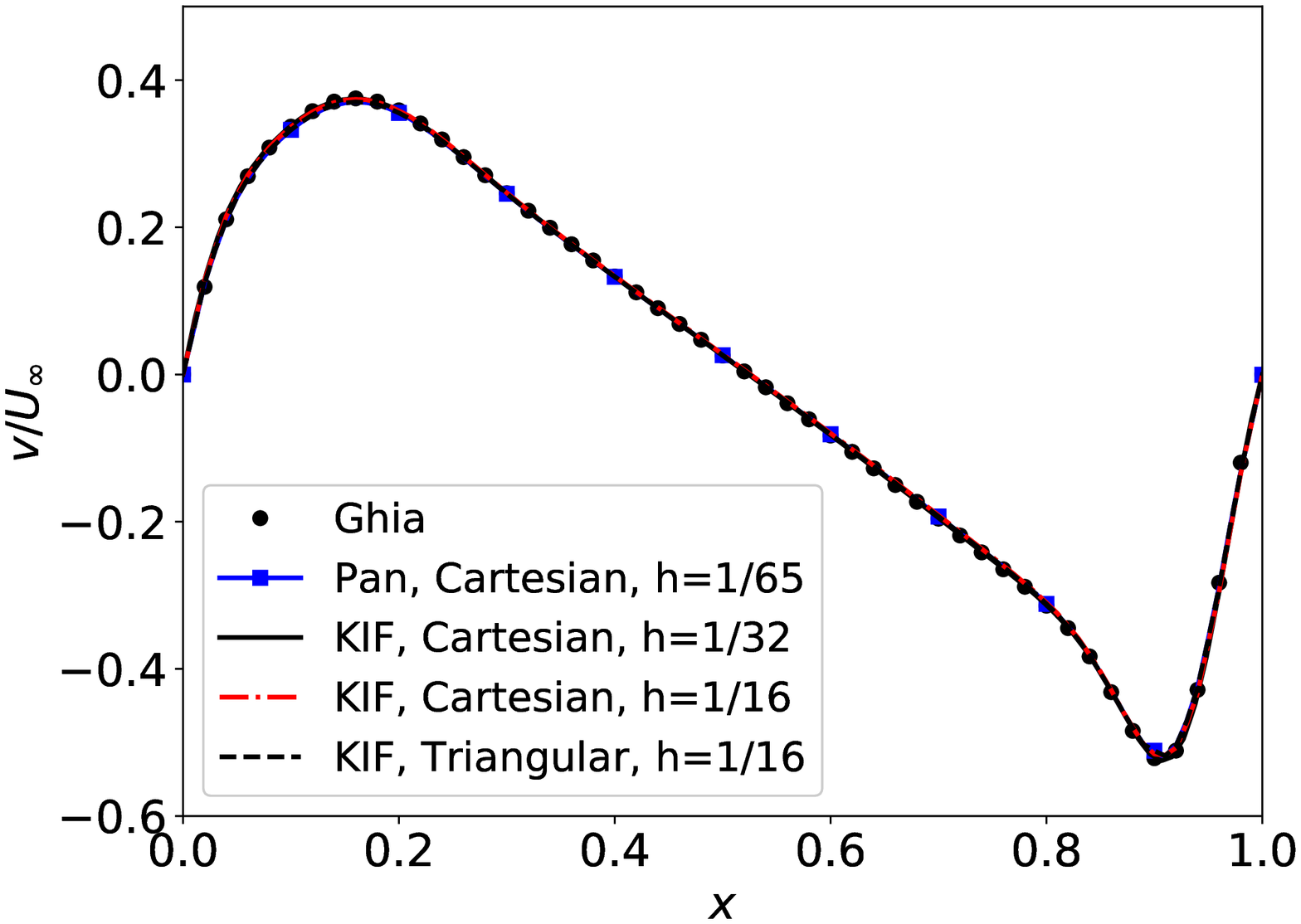}
		\label{fig:cavity:v:re1000}
	}
	\caption{The velocity profile of the lid driven cavity flow at $Re = 1000$: (a) u-velocity profiles at $x = 0.5$ and (b) v-velocity profiles at $y = 0.5$.}
	\label{fig:cavity:uv:re1000}
\end{figure}

\begin{figure}[!htp]
	\centering
	\subfigure[]{
		\includegraphics[width=0.4 \textwidth]{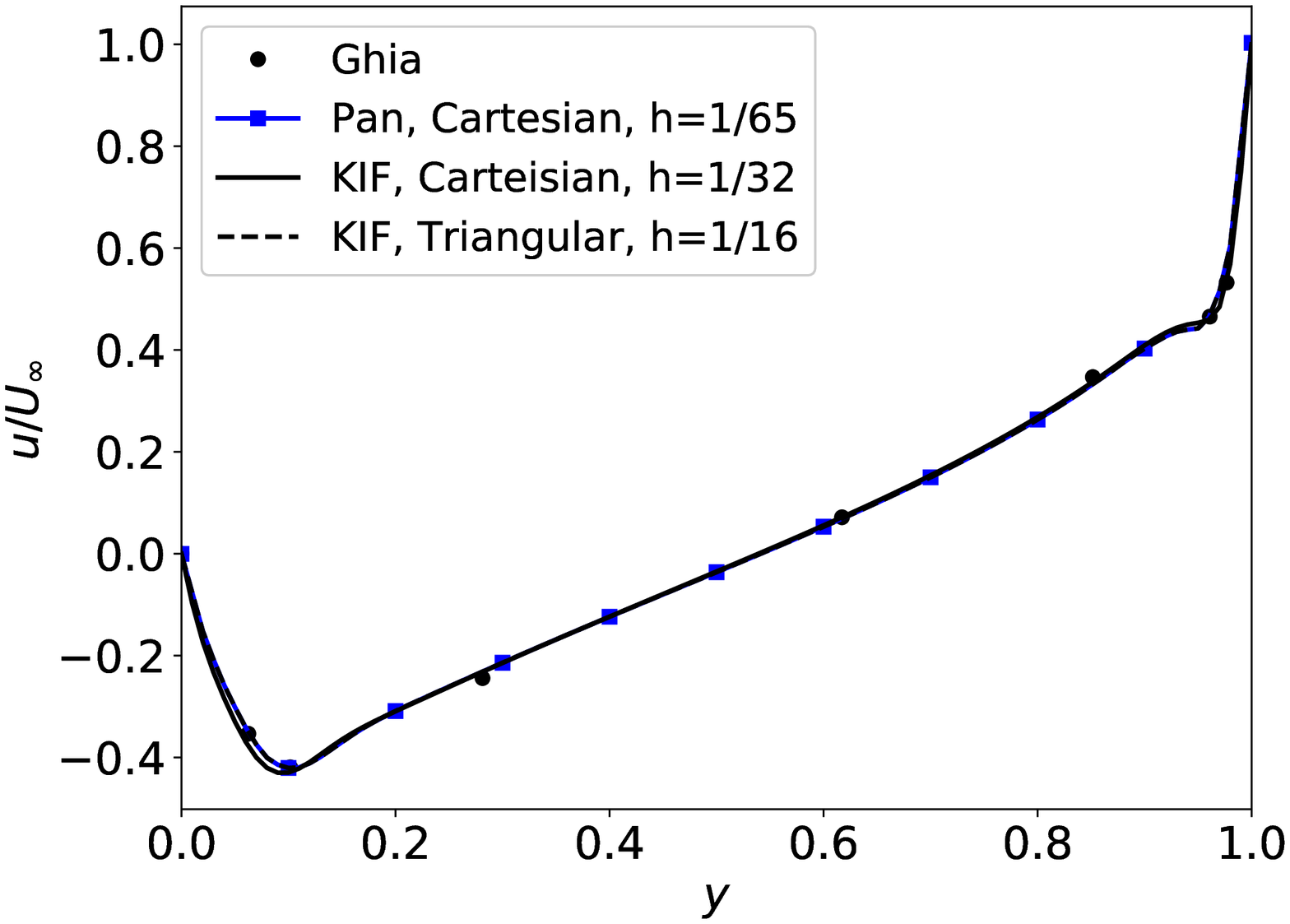}
		\label{fig:cavity:u:re3200}
	}
	\subfigure[]{
		\includegraphics[width=0.4 \textwidth]{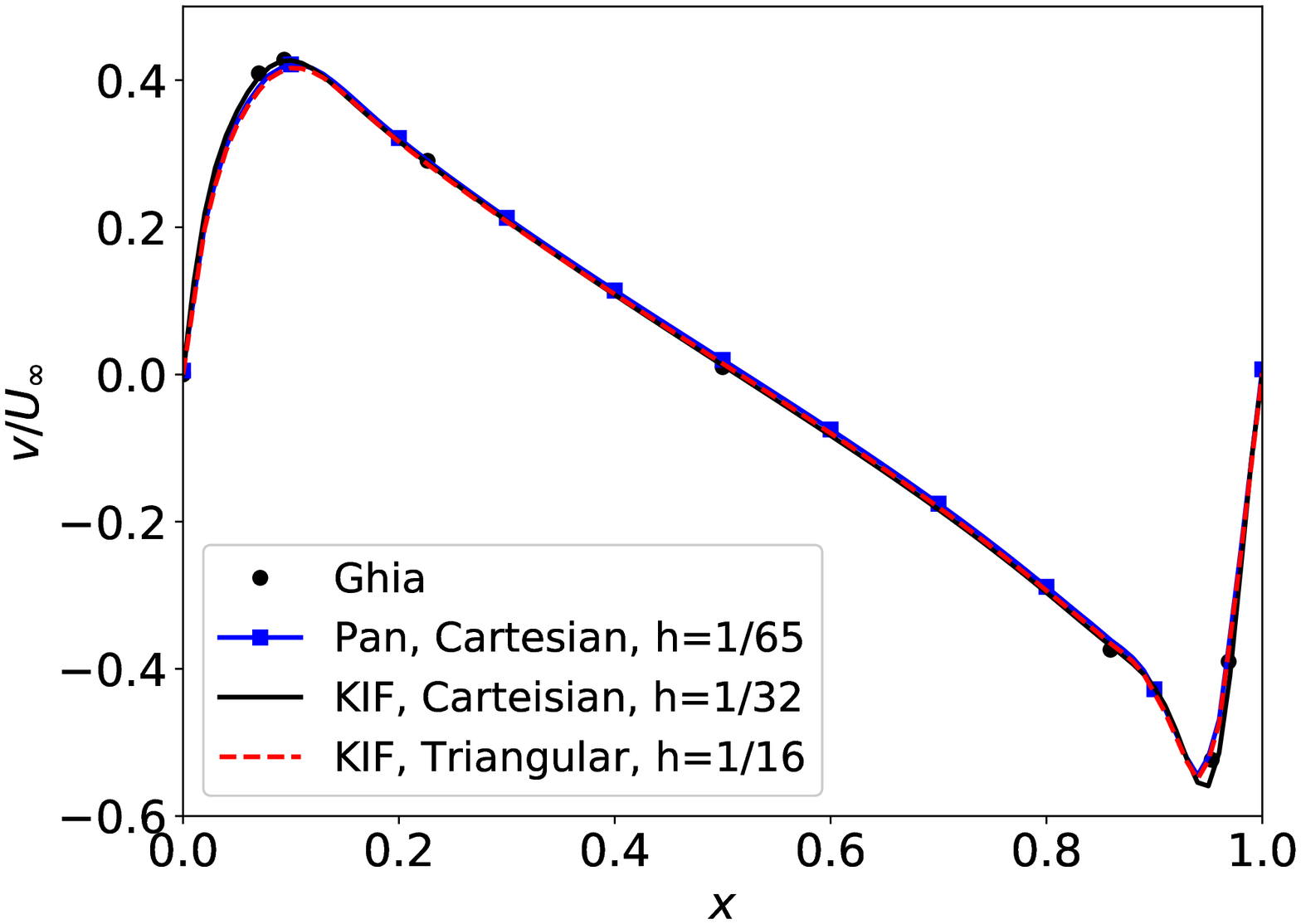}
		\label{fig:cavity:v:re3200}
	}
	\caption{The velocity profile of the lid driven cavity flow at $Re = 3200$: (a) u-velocity profiles at $x = 0.5$ and (b) v-velocity profiles at $y = 0.5$.}
	\label{fig:cavity:uv:re3200}
\end{figure}

The lid-driven cavity flow~\cite{ghia1982high} is one of the benchmarks for validating the performance of the viscous flow solver, and the aim of this case is also to examine performance of present method on viscous solid wall. An incompressible flow is initialed in the computational domain, and the Mach number of the lid is set as $Ma=0.1$. The Reynolds number are $Re = 400, 1000, 3200$ respectively. The computational domain is $[0,1]\times[0,1]$, and both the Cartesian and triangular grids are used in the computation. Fig.~\ref{fig:cavity:grid_tri} shows the triangular grid used in the case, and the length scale of the triangular grid is $h = 1/16$. The u-velocity profiles along the vertical center-line and the v-velocity profiles along the horizontal center-line are all compared with the reference data in Figs.~\ref{fig:cavity:uv:re400}-\ref{fig:cavity:uv:re3200}. The figures exhibit that the present results match quite well with the data from Ghia~\cite{ghia1982high}. The numerical data extracted from Ref.~\cite{pan2016efficient} at $Re=1000, 3200$ on $65 \times 65$ Cartesian grid are also shown in the figures. It is obvious that the present method can reach the same accuracy with fewer mesh nodes.

\subsection{Blasius incompressible laminar flat plate}
\begin{figure}[!htp]
	\centering
	\subfigure[]{
		\includegraphics[width=0.4 \textwidth]{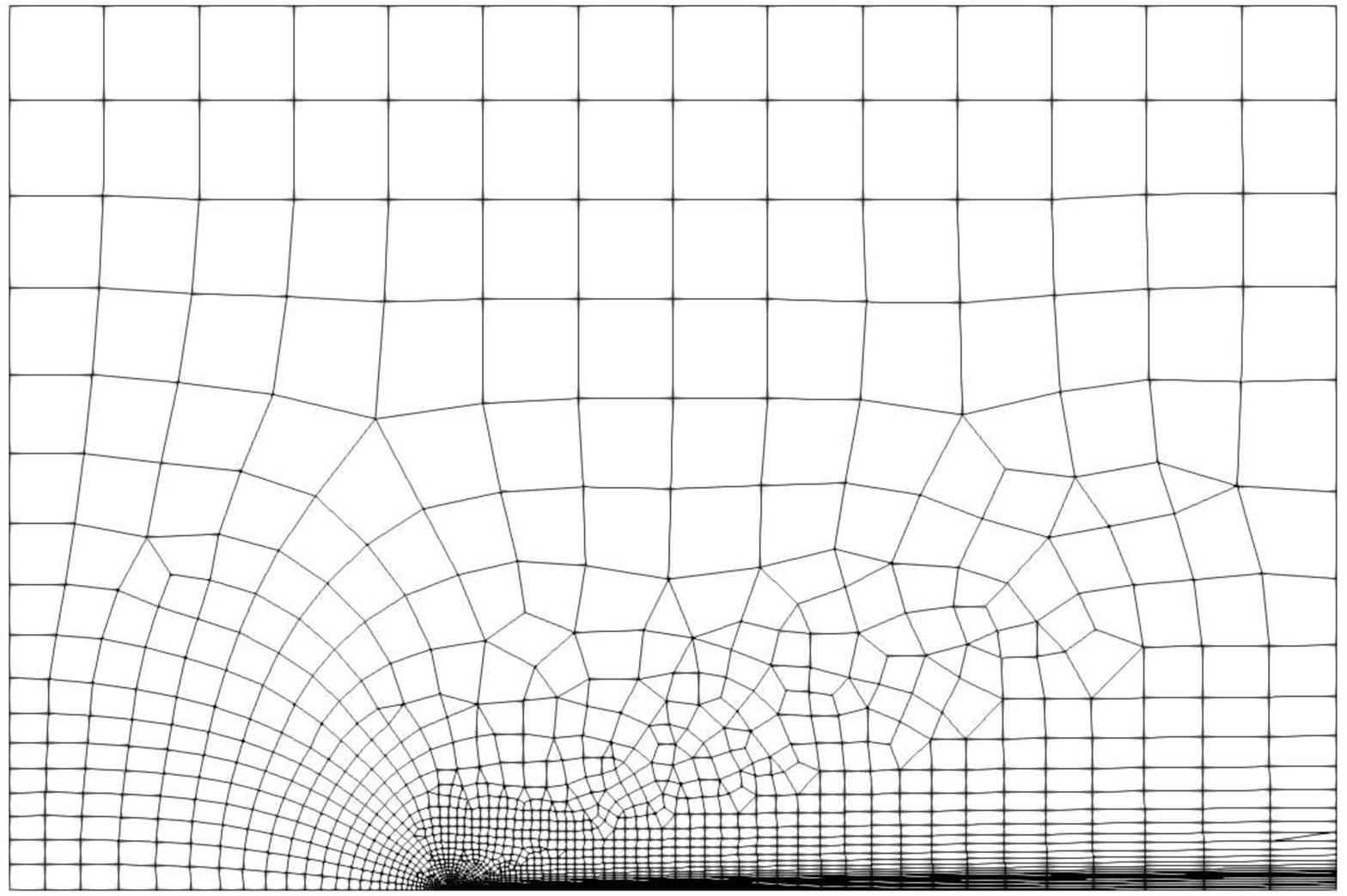}
		\label{fig:lam_flatplate:Grid2:1}
	}
	\subfigure[]{
		\includegraphics[width=0.4 \textwidth]{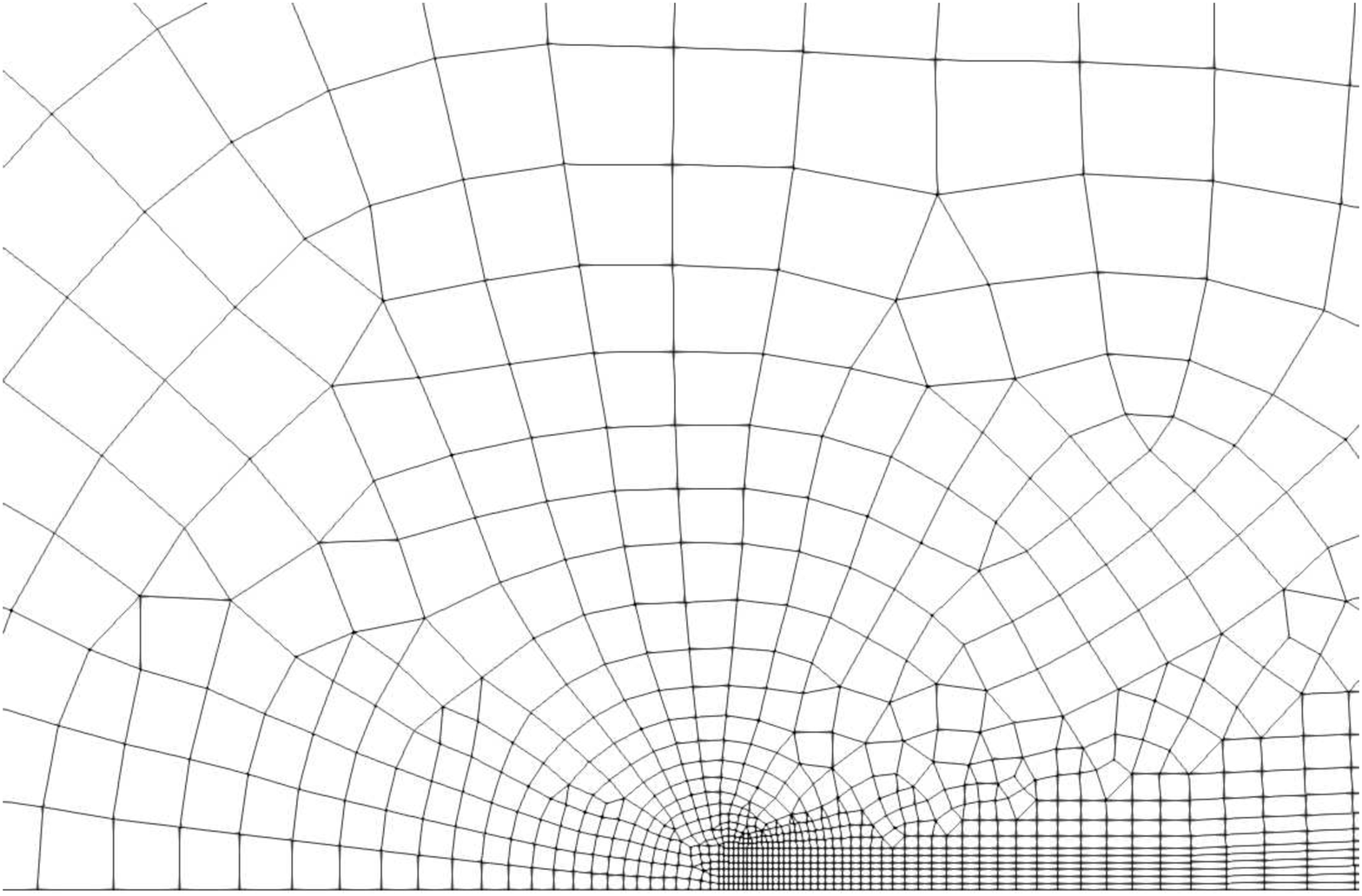}
		\label{fig:lam_flatplate:Grid2:2}
	}
	\caption{\label{fig:flatplate:grid2} The ``Grid 2" used in the simulation of laminar flow over a flat plate.}
\end{figure}

The incompressible boundary layer flow over a flat plate is simulated. The Mach number is $Ma_{\infty} = 0.15$, and the Reynolds number based on the length of plate is $Re_{\infty}=1\times10^5$. The subscript $\infty$ indicates the state of free stream flow.  The length of the flat plate is $L=100$, and the leading edge of the flat plate is located at $x=0$. The computational domain is $[-50, 100]\times[0,100]$. The inflow boundary is applied on the left side of the domain. The upper and right side of the domain is treated as subsonic outflow boundary. The viscous solid wall is used on the flat plate. The symmetric boundary is implemented at the bottom from the left side to the leading of the plate. It is known to all that using hybrid grid can reduce the grid size obviously, and the hybrid grid is much flexible than structured grid. To show the advantages of hybrid grid,  two hybrid grids are considered in the simulation. The coarser grid is shown in Fig.~\ref{fig:flatplate:grid2}, and the details of the two grids are shown in Table ~\ref{table:flatplate:grid}. ``Grid 2'' has fewer elements than ``Grid 1" in the boundary layer.

\begin{figure}[!htp]
	\centering
	\subfigure[]{
		\includegraphics[width=0.4 \textwidth]{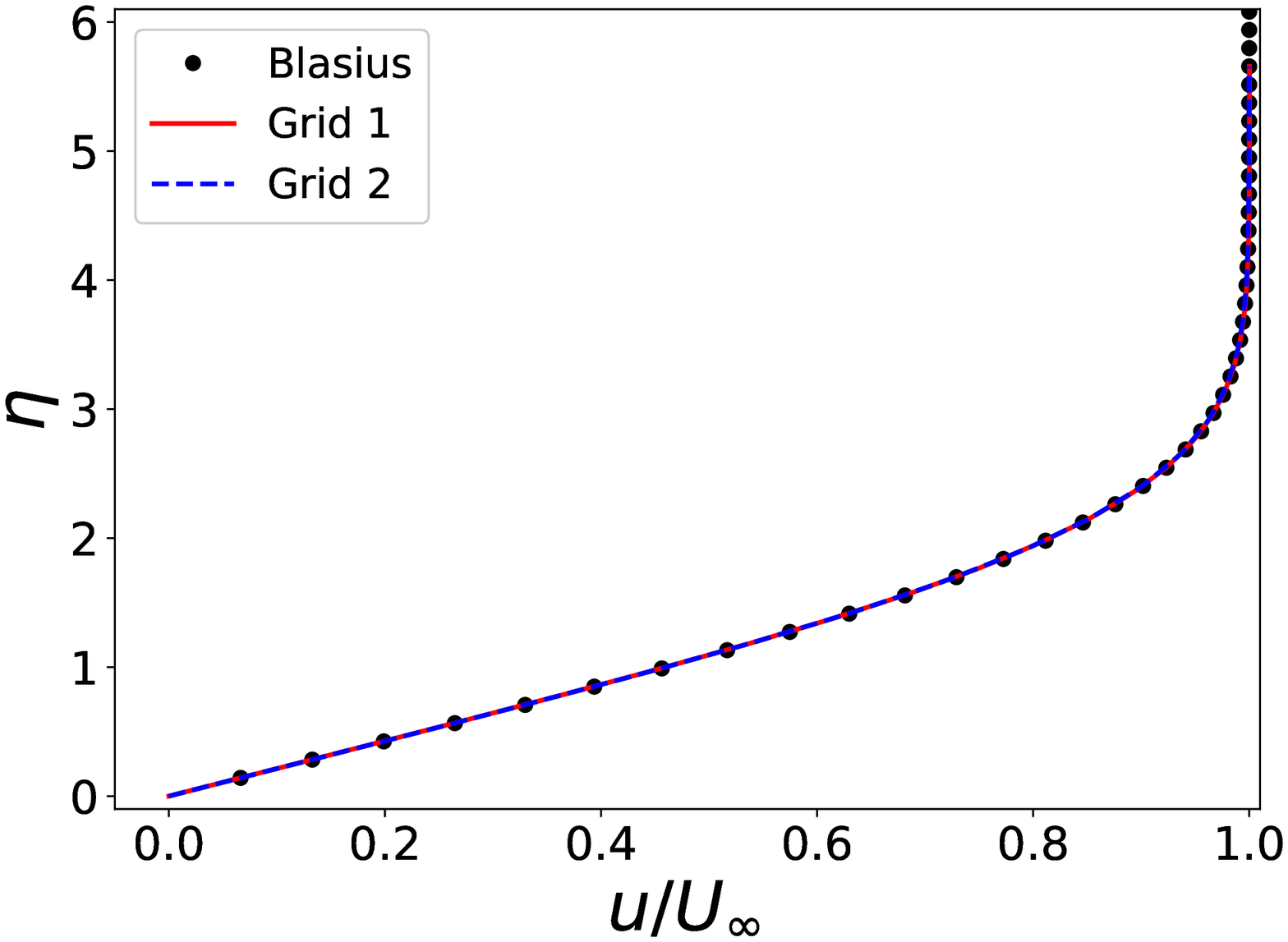}
		\label{fig:lam_flatplate:u:0.1}
	}
	\subfigure[]{
		\includegraphics[width=0.4 \textwidth]{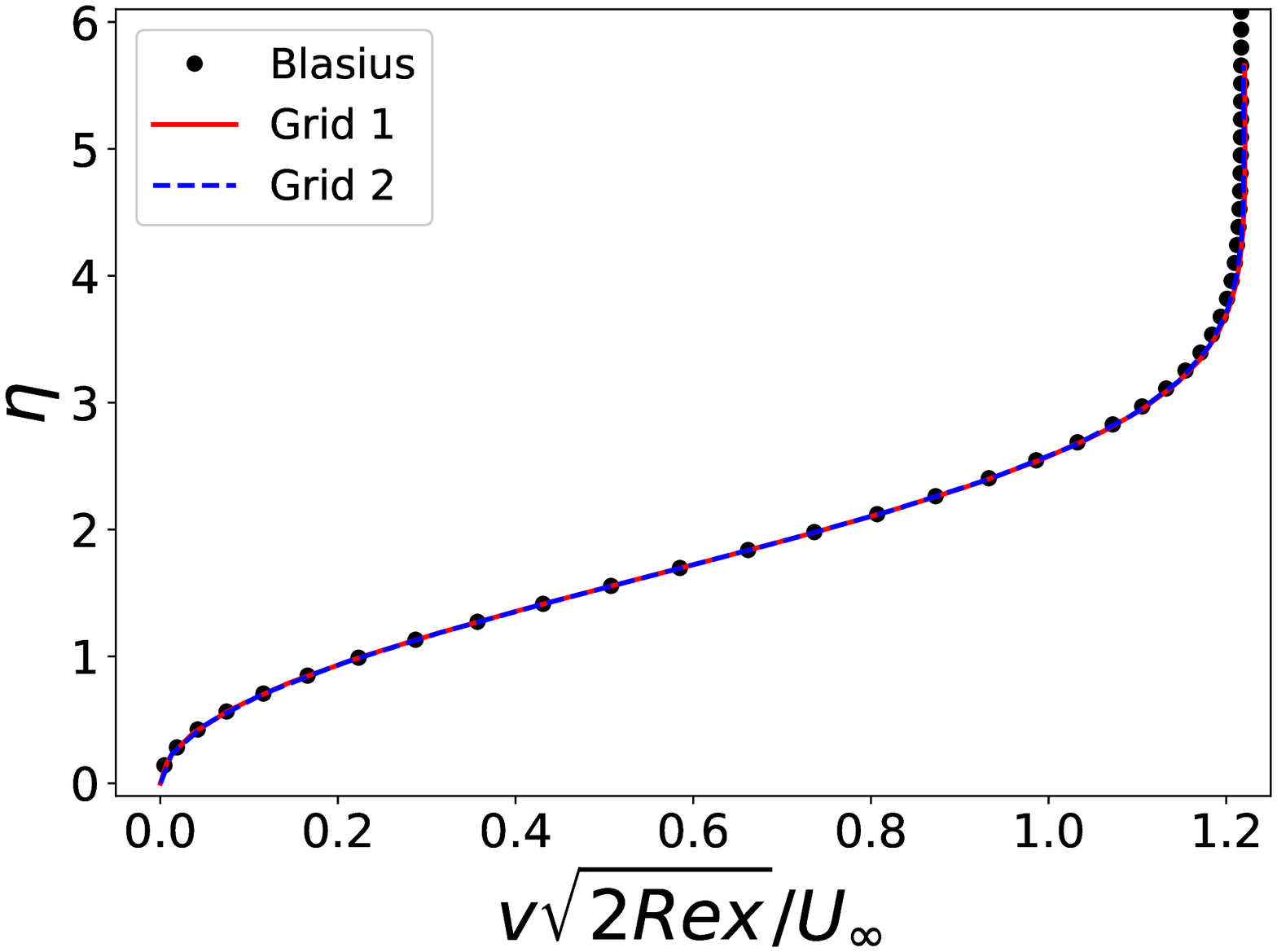}
		\label{fig:lam_flatplate:v:0.1}
	}
	\caption{Blasius incompressible laminar flat plate: (a) u-velocity and (b) v-velocity profiles at $x/L = 0.1$.}
	\label{fig:lam_flatplate:uv:0.1}
\end{figure}

\begin{figure}[!htp]
	\centering
	\subfigure[$u$ profile]{
		\includegraphics[width=0.4 \textwidth]{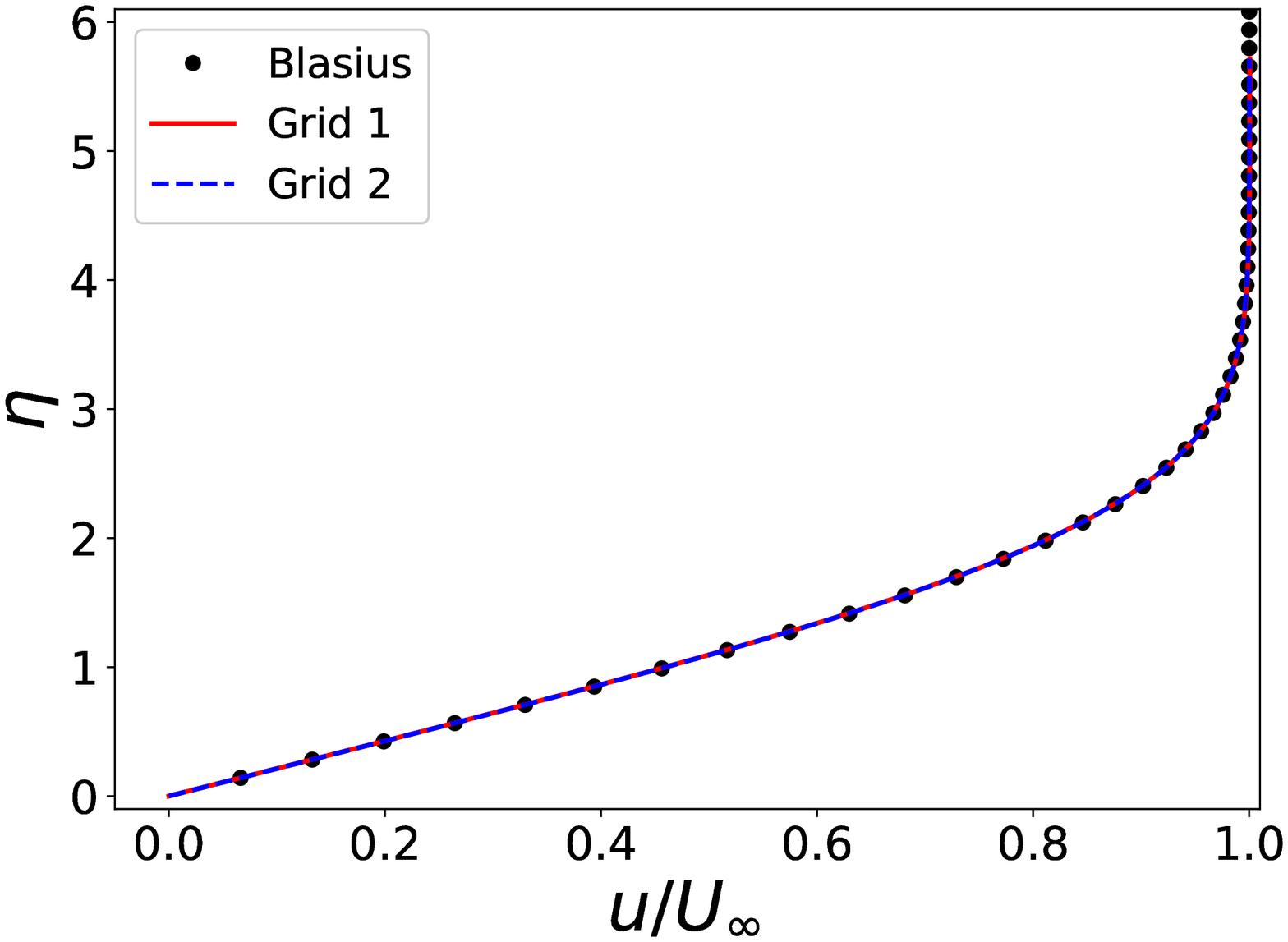}
		\label{fig:lam_flatplate:u:0.3}
	}
	\subfigure[$v$ profile]{
		\includegraphics[width=0.4 \textwidth]{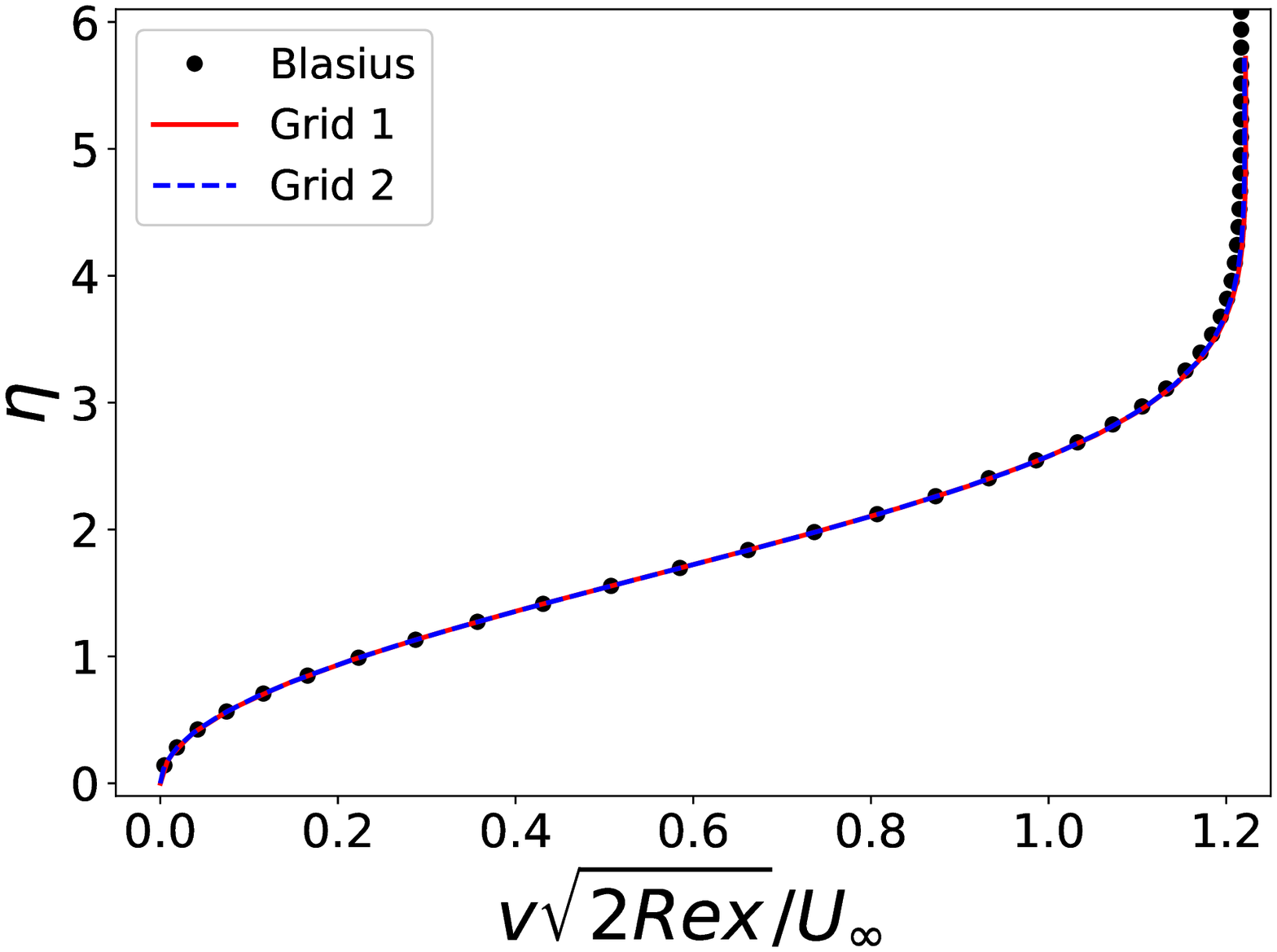}
		\label{fig:lam_flatplate:v:0.3}
	}
	\caption{Blasius incompressible laminar flat plate: (a) u-velocity and (b) v-velocity profiles at $x/L = 0.3$.}
	\label{fig:lam_flatplate:uv:0.3}
\end{figure}

\begin{figure}[!htp]
	\centering
	\subfigure[$u$ profile]{
		\includegraphics[width=0.4 \textwidth]{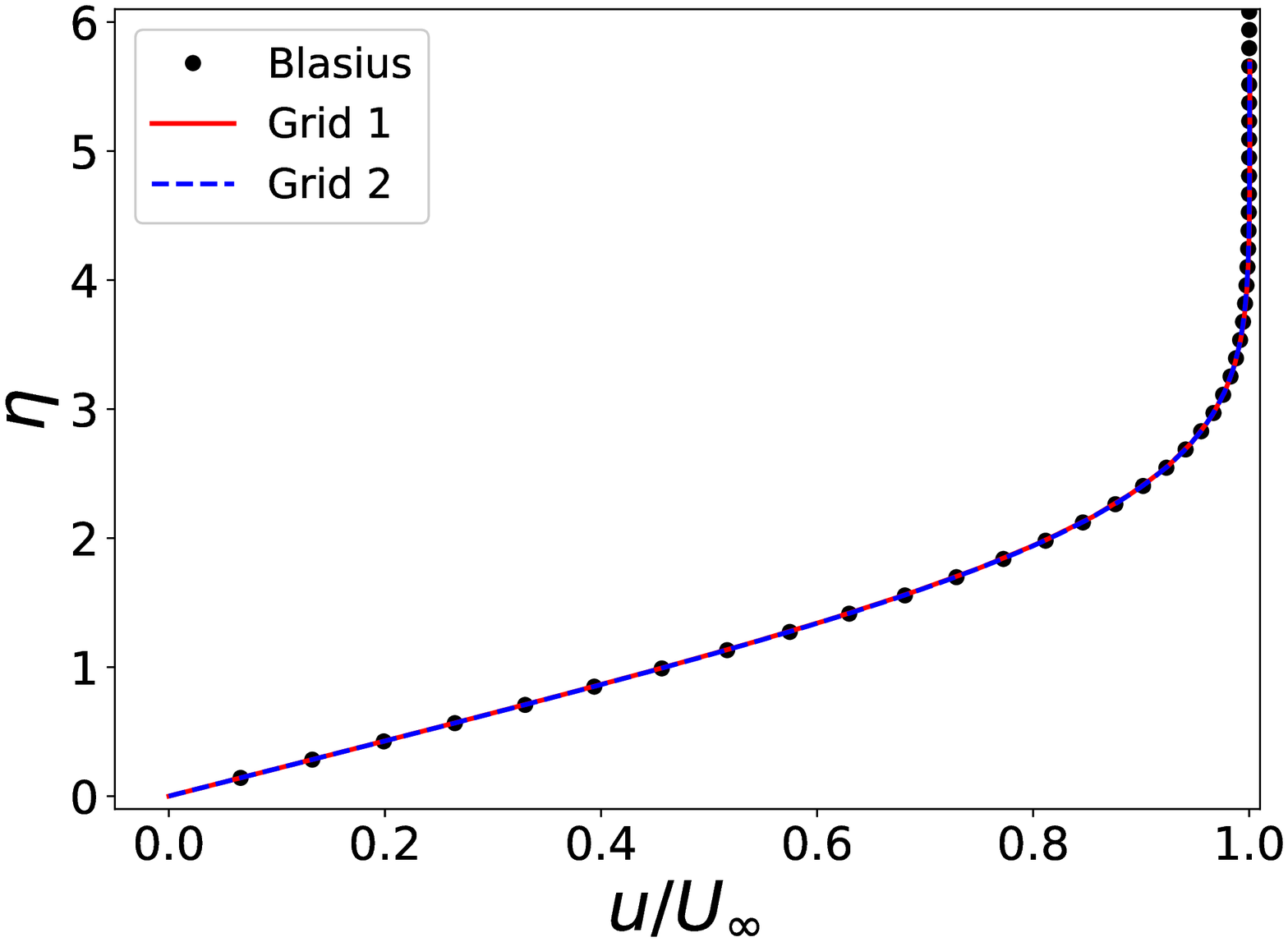}
		\label{fig:lam_flatplate:u:0.5}
	}
	\subfigure[$v$ profile]{
		\includegraphics[width=0.4 \textwidth]{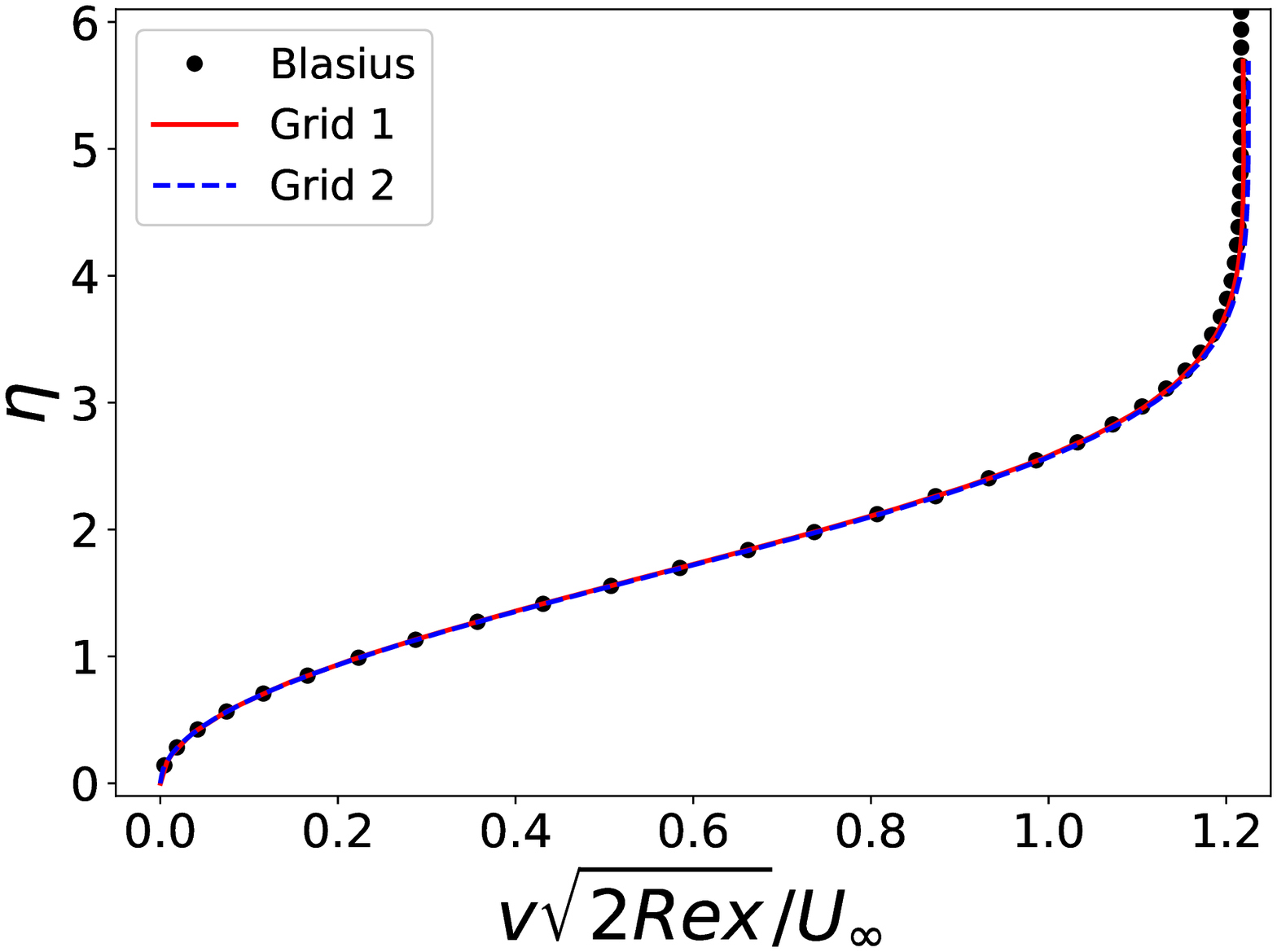}
		\label{fig:lam_flatplate:v:0.5}
	}
	\caption{Blasius incompressible laminar flat plate: (a) u-velocity and (b) v-velocity profiles at $x/L = 0.5$.}
	\label{fig:lam_flatplate:uv:0.5}
\end{figure}

\begin{figure}[!htp]
	\centering
	\includegraphics[width=0.4 \textwidth]{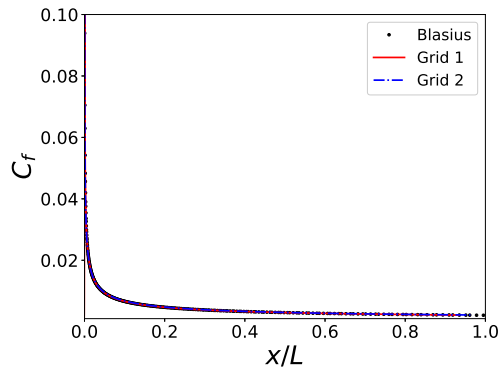}
	\caption{\label{fig:flatplate:cf} Blasius incompressible laminar flat plate: the skin friction coefficient distribution along the flat plate.}
\end{figure}

The velocity profiles versus $\eta=y\sqrt{\frac{U_{\infty}}{\nu x}}$ are shown in Figs.~\ref{fig:lam_flatplate:uv:0.1}-\ref{fig:lam_flatplate:uv:0.5}. It can be seen clearly in the pictures that the velocity profiles have a good accordance with the Blasius solution at every positions. It is obvious that the present method can approach boundary layer with very few elements. The skin friction coefficient is shown in Fig.~\ref{fig:flatplate:cf}. The numerical results shows an excellent performance compared with Blasius solution.

\begin{table}[!htp]
	\centering
	\caption{\label{table:flatplate:grid} The details of the grids used in the laminar boundary layer case.}
	\begin{threeparttable}
		\begin{tabular}{p{40pt} p{100pt}< {\centering} p{100pt} < {\centering}}
			\hline
			\hline
			 & Grid size &  Number of elements in boundary layer \\
			\hline
			Grid 1 & $4876$ & $25$ \\
			
			Grid 2 & $2336$ & $6$ \\
			
			\lasthline
		\end{tabular}
	\end{threeparttable}
\end{table}

\subsection{Taylor-Green vortex at $Re = 1,600$}
\begin{figure}[!htp]
 	\centering
 	\subfigure[]{
 		\includegraphics[width=0.4 \textwidth]{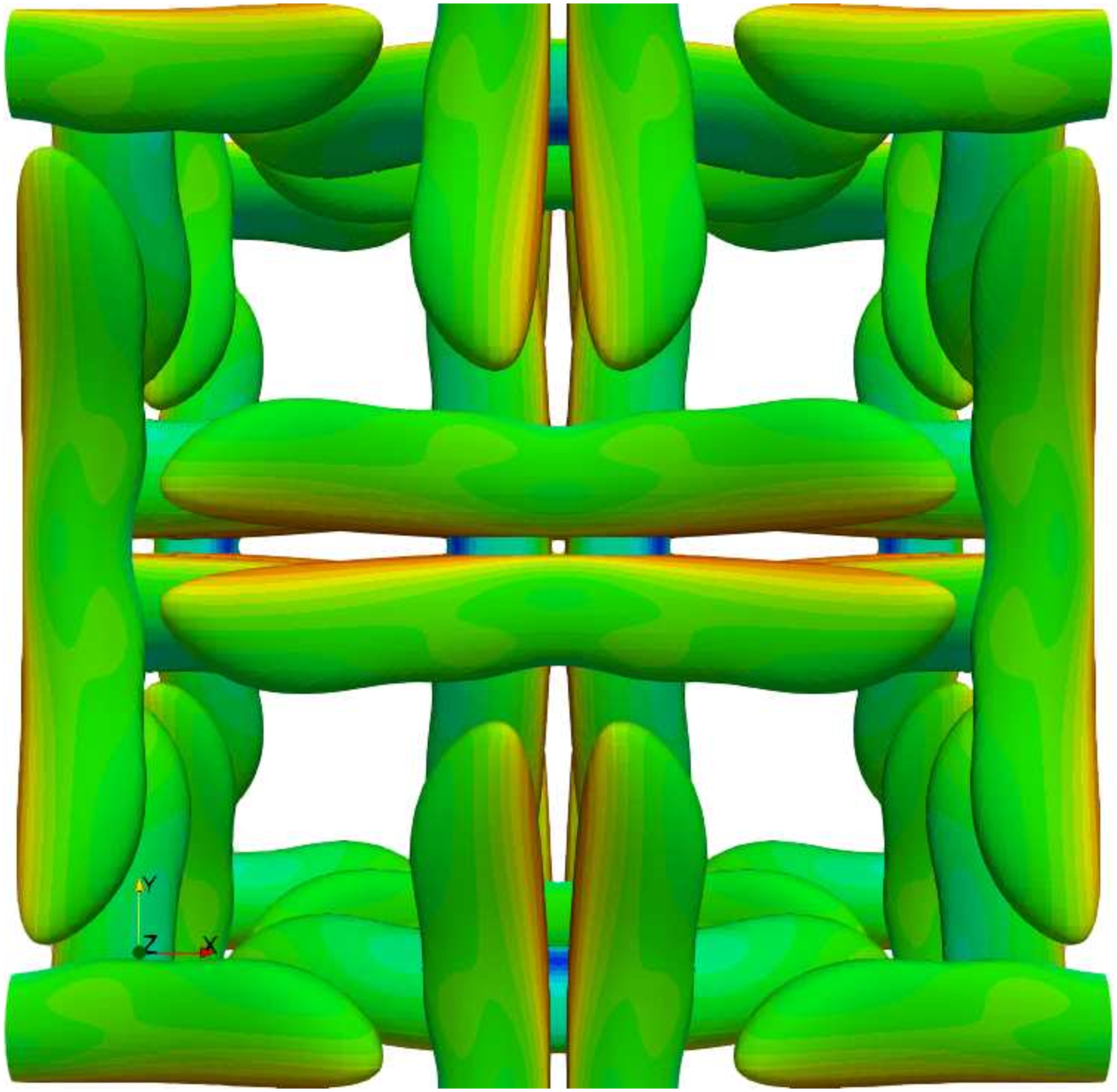}
 		\label{fig:TGV:Q:t:3}
 	}
 	\subfigure[]{
 		\includegraphics[width=0.4 \textwidth]{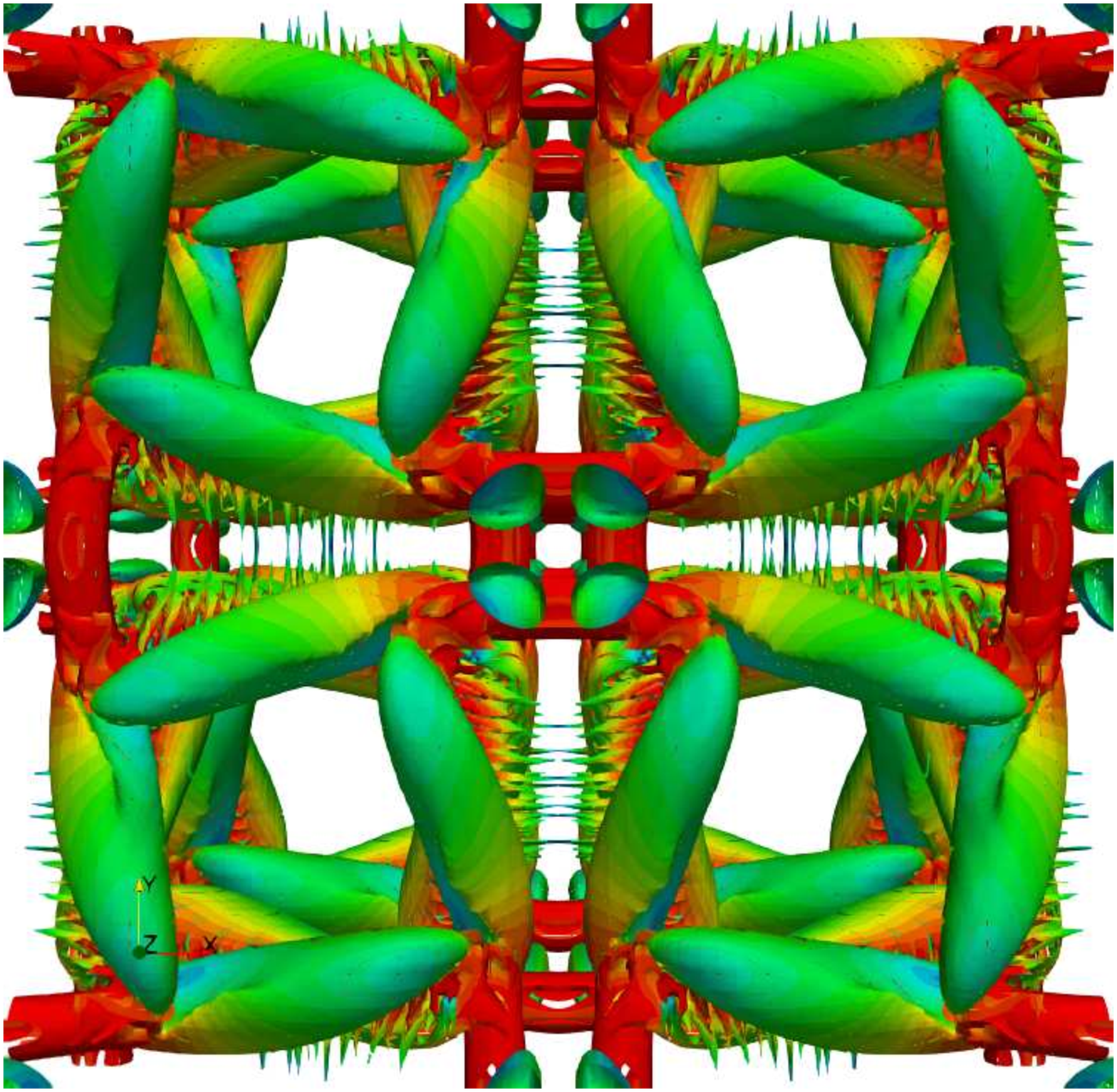}
 		\label{fig:TGV:Q:t:5}
 	}
 	\subfigure[]{
		\includegraphics[width=0.4 \textwidth]{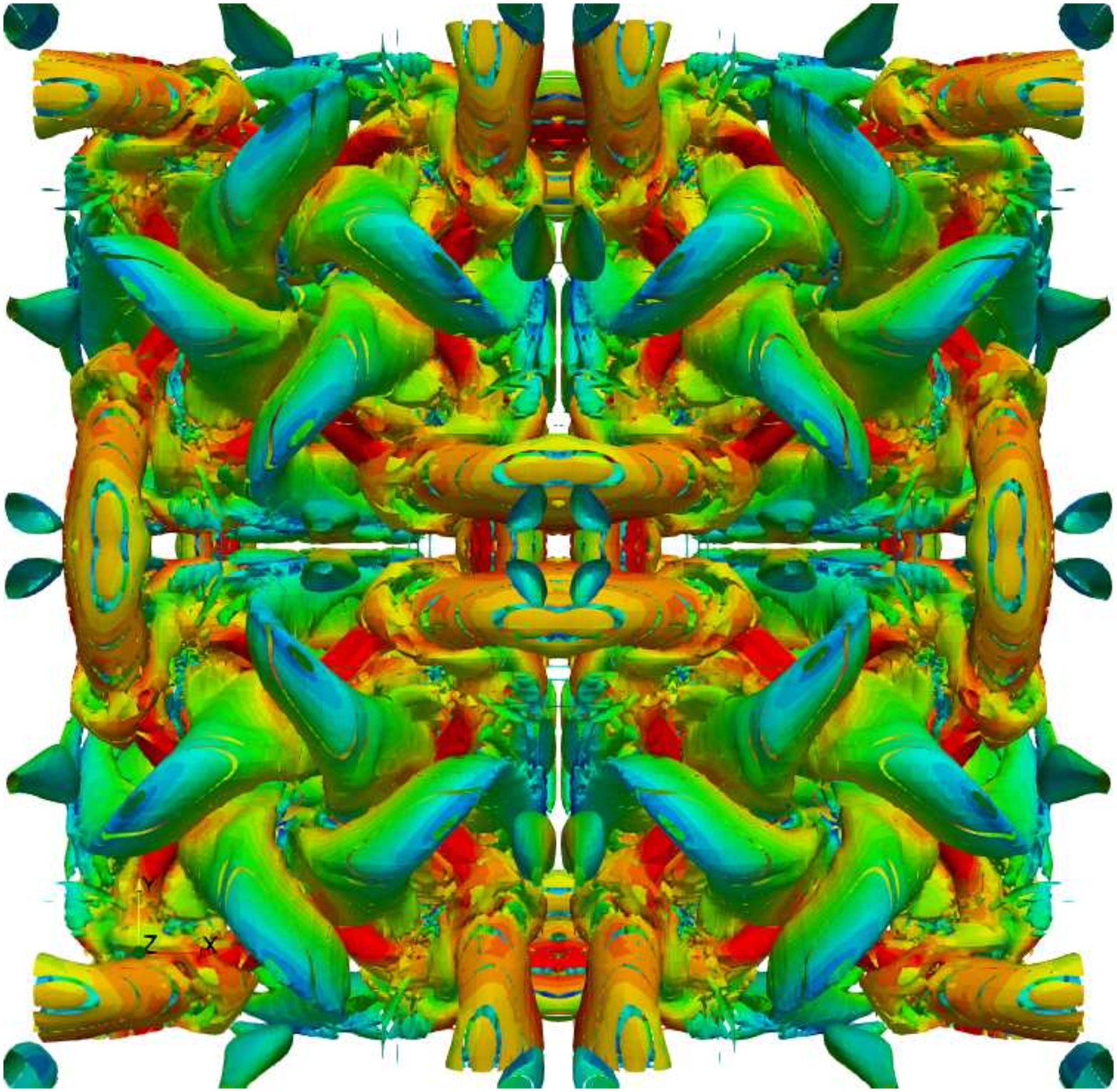}
		\label{fig:TGV:Q:t:7}
	}
	\subfigure[]{
		\includegraphics[width=0.4 \textwidth]{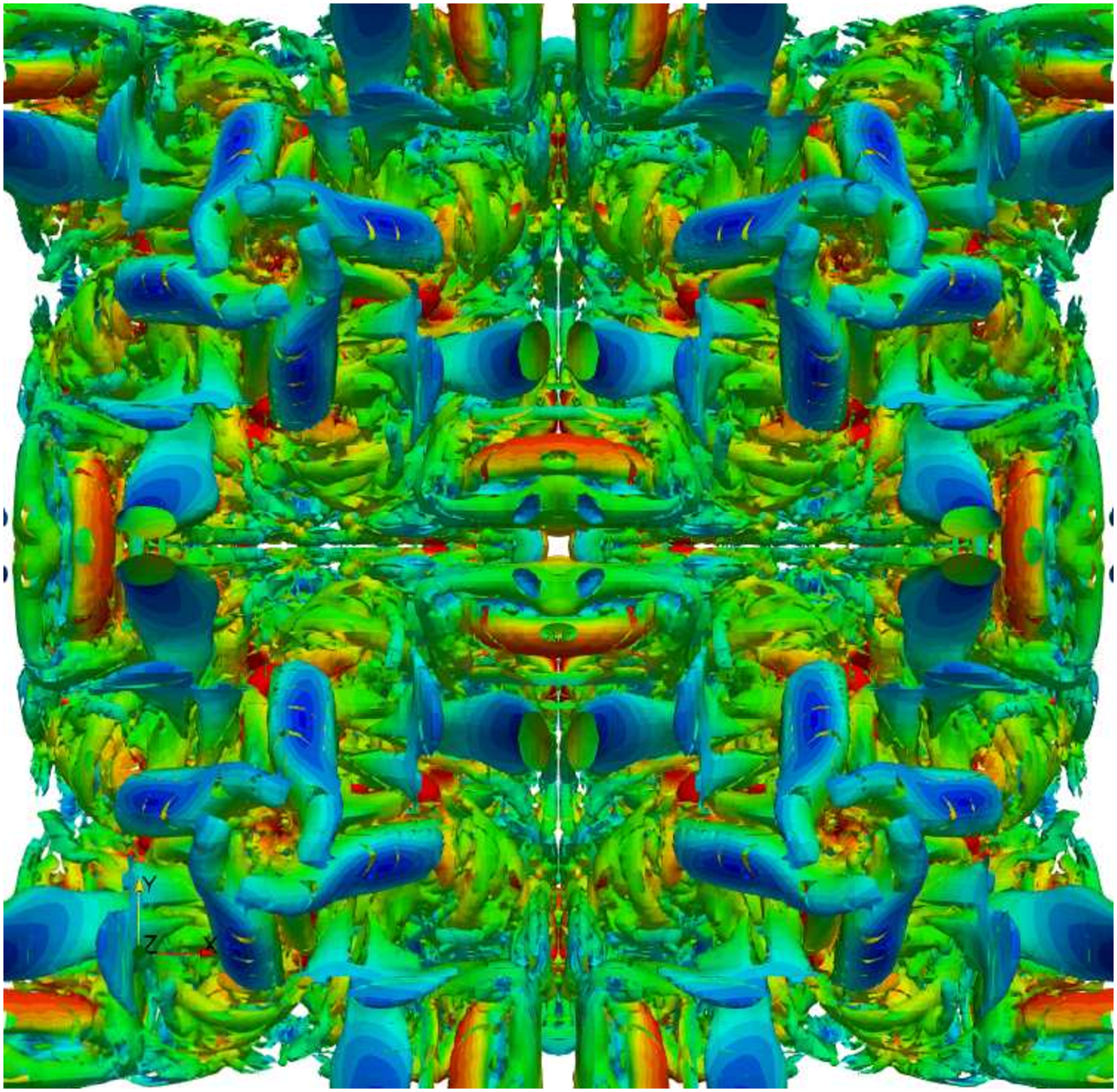}
		\label{fig:TGV:Q:t:9}
	}
 	\caption{\label{fig:TGV:Q} Iso-surfaces of Q ($Q=0.5$) criterion colored by velocity magnitude: (a) $t = 3$, and
 		(b) $t = 5$, (c) $t = 7$, and (d) $t = 9$.
 	}
 	
 \end{figure}

The Taylor-Green Vortex (TGV) is a simple test case for the resolution of the small scales of a turbulent flow by a numerical method. The compressible TGV at $Re = 1600$ was one of the benchmark problems in the 1st and 2nd International Workshops on High-Order CFD Methods. The reference solution used in current paper was obtained by Debonis~\cite{debonis_solutions_nodate} using a high-order dispersion-relation-preserving (DRP) scheme on a mesh of $512^3$ elements. The computational domain is a cubic box of dimensions $[0, 2\pi]^3$, and the periodic boundary is applied on the faces of the cube. In the case, both of the $32^3$ and $64^3$ grids are used.\\
The initial condition is set as

\begin{equation}\label{TGV3D:init}
\begin{aligned}
u(t_0)&=u_0sin(x/L)cos(y/L)cos(z/L),\\
v(t_0)&=-u_0cos(x/L)sin(y/L)cos(z/L),\\
w(t_0)&=0, \\
p(t_0)&=p_0+\frac{\rho_0V_0^2}{16}\left[cos(2x/L)+cos(2y/L)\right]\left[cos(2z/L)+2\right],
\end{aligned}
\end{equation}
where $\rho_0 = 1$, $p_0 = 100$, $u_0 = 1$, and $L = 1$. The Mach number is set to $0.08$, and the initial temperature is $300K$. The volume-averaged kinetic energy and the dissipation rate of the kinetic energy are computed. The volume-averaged kinetic energy is read as
\begin{equation}\label{KineticEnergy}
E_k = <k> =\frac{1}{\rho_0\Omega}\int_{\Omega}\rho \frac{u_iu_i}{2}d\Omega,
\end{equation}
and the dissipation rate of the kinetic energy is give by
\begin{equation}\label{KEDR_epsilon}
\epsilon(E_k) =-\frac{dE_k}{dt}.
\end{equation}
The numerical results of averaged kinetic energy and the dissipation rate of kinetic energy are compared with reference data from Debonis~\cite{debonis_solutions_nodate}. The results have a good accordance with the reference data. The iso-surfaces of Q criterions colored by velocity magnitude at time $3$, $5$, $7$, and $9$ are shown in Fig.~\ref{fig:TGV:Q}. The evolution of flow structure from large scale vortices to small vortices can be clearly seen in the figure.

\begin{figure}[!htp]
	\centering
	\subfigure[]{
		\includegraphics[width=0.4 \textwidth]{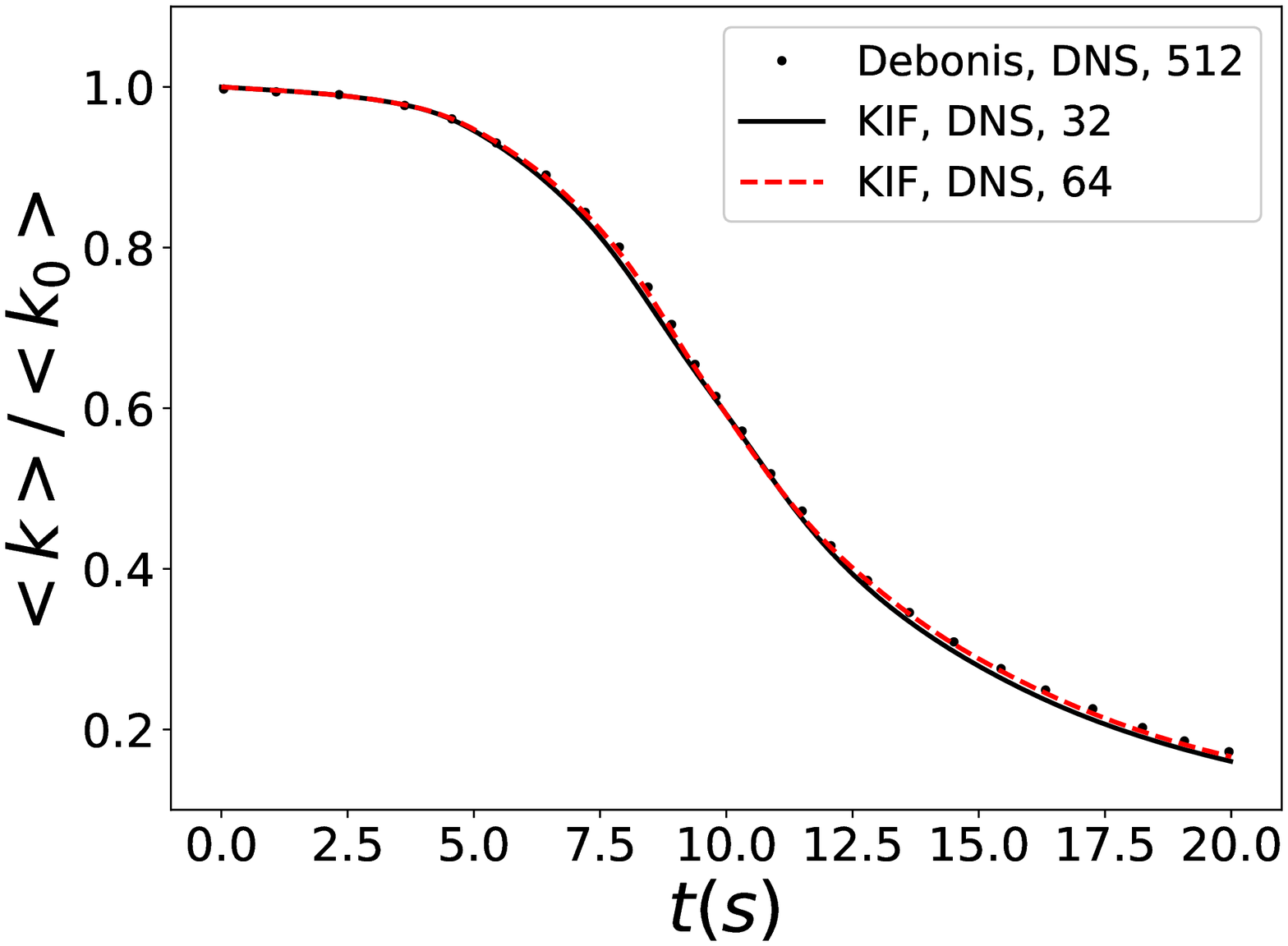}
		\label{fig:TGV:epsilon_1}
	}
	\subfigure[]{
		\includegraphics[width=0.4 \textwidth]{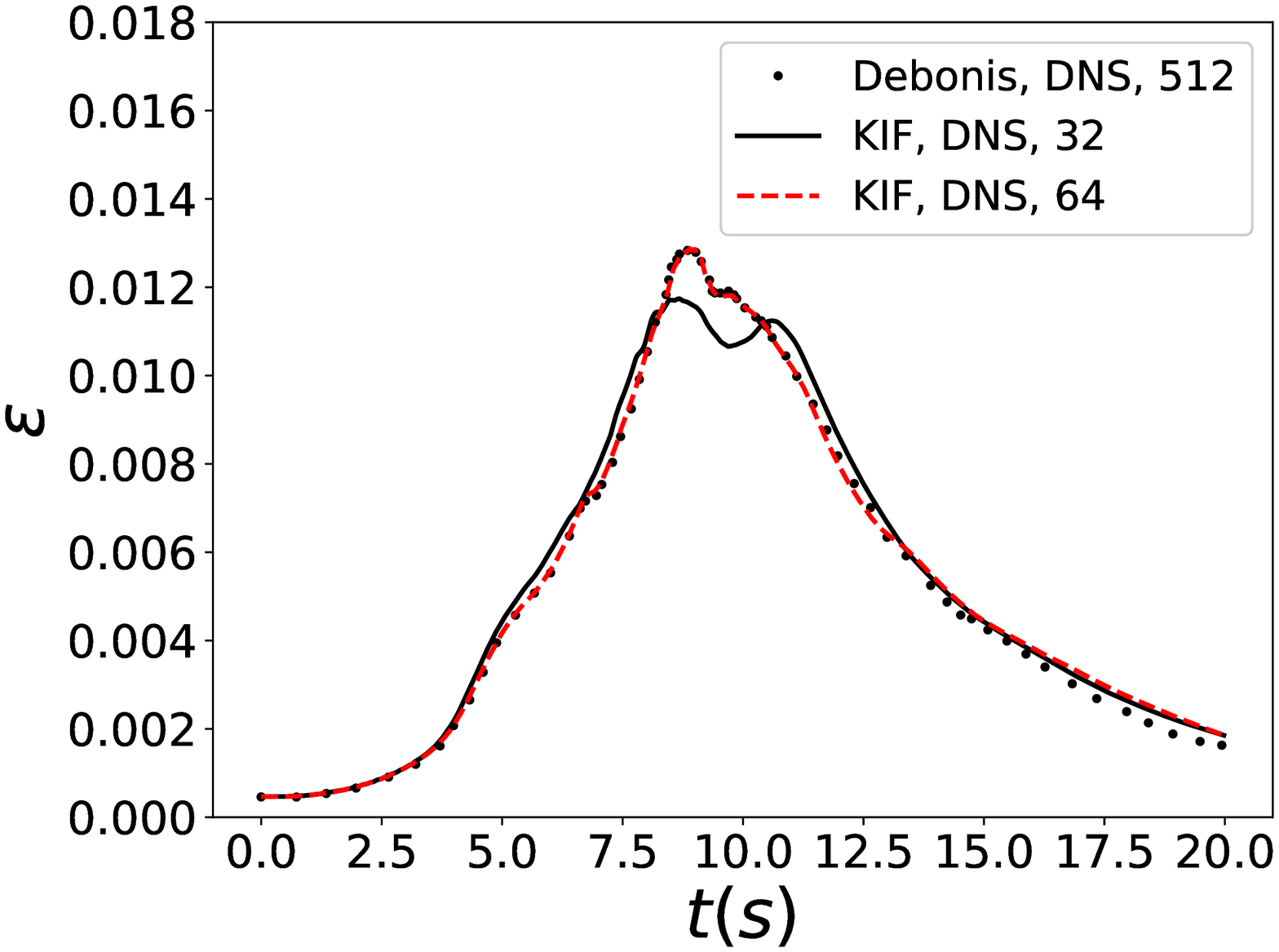}
		\label{fig:TGV:epsilon_2}
	}
	\caption{The kinetic energy and the dissipation rate: (a) $E_k$ and (b) $\epsilon(E_k)$. The integer number denotes the grid size. For example, the number $32$ indicates that the grid size is $32^3$.}
	\label{fig:TGV:KDER}
\end{figure}

\section{Conclusion}\label{conclusion}
In present paper, a high order numerical scheme is proposed based on the flux reconstruction framework and Kinetic Inviscid Flux. KIF, which aims to find a balance between the advantages of gas-kinetic scheme and lower computational costs, is a combination of TTT scheme and KFVS method. FR framework is well understood and available. It has the properties of robustness and compactness, which are of great importance for the high order method. The accuracy order have been verified using the advection of density perturbation problem and isentropic vortex problem. The results show that the accuracy of present method reaches to the designed order. KIF also can be viewed as an inviscid-viscous splitting version of the gas-kinetic scheme, and the excellent performance of proposed method can be seen in the simulations of lid-driven cavity flow and Blasius incompressible laminar boundary layer. The good numerical results have shown the success of inviscid-viscous strategy of gas-kinetic scheme. The Taylor-Green Vortex problem has been used to verify the potential of present method to simulate turbulent flow, and the excellent results are obtained. The shock capturing method used in current paper is still an open question which is needed in further study. The ability to capture strong discontinuity should be improved further.

\begin{acknowledgments}
The present work is supported by National Natural Science Foundation of China (Grant No. 11702223, No. 11902266 and No. 11902264), National Numerical Wind Tunnel Project of China, and 111 Project of China (Grant No. B17037), as well as the ATCFD Project (2015-F-016).
\end{acknowledgments}

\clearpage
\section*{References}
\bibliography{ref}
\clearpage
\listoffigures
\clearpage
\listoftables
\clearpage

\end{document}